**Photodynamic, UV-curable and fibre-forming polyvinyl alcohol derivative with broad processability and staining-free antibacterial capability**

Man Li,[1,2] Charles Brooker,[1,2] Rucha Ambike,[1] Ziyu Gao,[2,3] Paul Thornton,[1,3] Thuy Do,[2] Giuseppe Tronci[1,2,*]

[1] Clothworkers' Centre for Textile Materials Innovation for Healthcare, Leeds Institute of Textiles and Colour, School of Design, University of Leeds, Leeds, LS2 9JT, United Kingdom

[2] School of Dentistry, St. James's University Hospital, University of Leeds, Leeds, LS9 7TF, United Kingdom

[3] School of Chemistry, University of Leeds, Leeds, LS9 7TF, United Kingdom

## Abstract

Antimicrobial photodynamic therapy (APDT) is a promising antibiotic-free strategy for broad-spectrum infection control in chronic wounds, minimising bacterial resistance risks. However, rapid photosensitiser diffusion, tissue staining, side toxicity, and short-lived antimicrobial effects present significant clinical limitations for integrating APDT into wound dressings. To address these challenges, we present the design of a bespoke polyvinyl alcohol (PVA) derivative conjugated with both phenothiazine and methacrylate functionalities, enabling staining-free antibacterial photodynamic effects, cellular tolerability and processability into various wound dressing formats, including films, textile fibres and nanoscale coatings. Tosylation of PVA is leveraged for the covalent coupling of toluidine blue ([TB]: 0.69±0.03–0.81±0.05 mg per gram of polymer), as confirmed by UV-Vis spectroscopy and the minimal average release of TB (≤ 3 wt.%, < 0.4 μg) following 96-hour incubation *in vitro*. UV-induced network formation is demonstrated by complete solution gelation, rheology, and a high gel content ($\bar{G}$ > 95 wt.%), and exploited to accomplish cast films and nanoscale integrated wound dressing coatings. UV curing is also successfully coupled with an in-house wet spinning process to realise individual, water-insoluble fibres as the building blocks of fibrous wound dressings. A fluorometric assay supports the generation of reactive oxygen species when the UV-cured samples are exposed to work, but not UV, light, yielding a mean $\log_{10}$ reduction of up to 2.13 in *S. aureus*, and the complete eradication of *P. aeruginosa*. Direct and extract cytotoxicity tests with UV-cured films and fibres demonstrate the viability of L929 fibroblasts following 60-min light irradiation and 72-hour cell culture. The bespoke molecular architecture, broad processability and cellular tolerability of this PVA derivative are highly attractive aiming to integrate durable staining-free photodynamic capability in a wide range of healthcare technologies, from chronic wound dressings up to minimally invasive localised therapy.

* Corresponding author: g.tronci@leeds.ac.uk.

## Graphical abstract

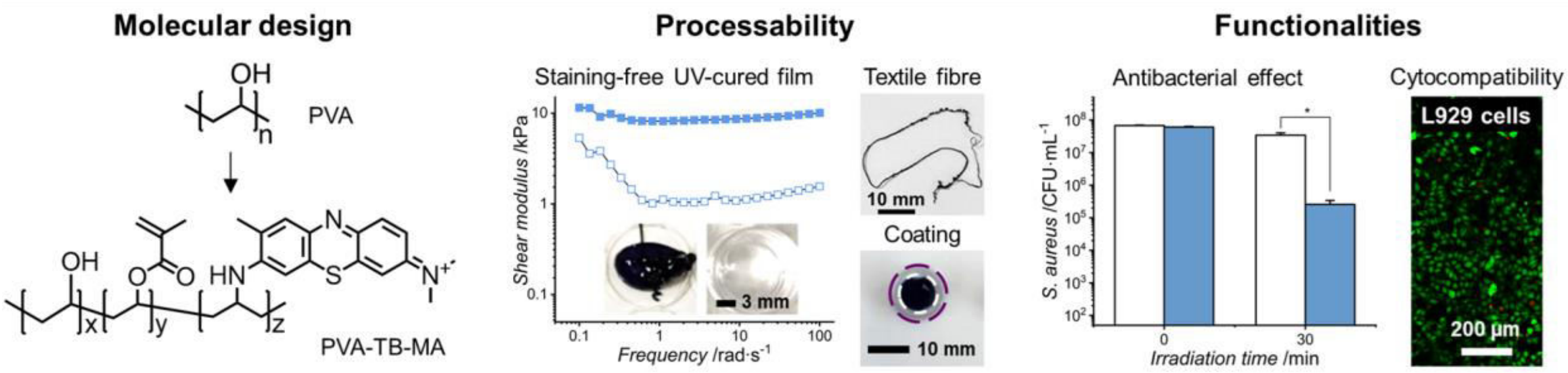

## Highlights

- A polyvinyl alcohol derivative with staining-free photodynamic effect is obtained
- Reactive oxygen species achieve up to complete bacterial killing *in vitro*
- Antibacterial photodynamic effect is enabled by visible, but not, UV light
- The polymer can be delivered as UV-cured film, textile fibre and nanoscale coating
- High cellular tolerability is observed *in vitro* in both extract and contact tests

## 1. Introduction

Wound management poses a significant economic burden worldwide. In the United

Kingdom alone, it costs the National Health Service (NHS) over £8 billion annually, with approximately two-thirds allocated to managing unhealed wounds [1]. Chronic wounds fail to progress through the normal stage of wound healing and usually take more than three months to heal. Prolonged exposure to exogenous pathogens, such as bacteria, significantly increases the risks of wound infection and irreversible complications, including gangrene and amputation [2,3,4]. Effective infection control in wound dressings is crucial for accelerating healing, reducing pain, and minimising infection recurrence.

Treating wound infections is increasingly challenging due to rising bacterial resistance and delayed infection diagnosis [5,6,7], ultimately causing further healing delays, prolonged hospitalisation time, and unaffordable healthcare costs. There is, therefore, an urgent need for antibiotic-free infection control strategies with broad-spectrum activity that can fight infection with minimal risks of generating antimicrobial resistance [8]. These strategies should be developed to enable scalability and integration with medical devices, i.e. wound dressings, to foster localised antimicrobial action and minimise negative effects on the wound environment and surrounding tissues and cells [9]. Ultimately, they should also be patient-friendly to support the ongoing shift towards home-based care, to reduce the need for continuous hospital visits, saving nursing time and healthcare costs.

APDT has emerged as an efficient, non-invasive antibiotic-free antimicrobial strategy for accessible tissues, enabling precise temporal and spatial control alongside broad-spectrum antimicrobial effects [10,11]. APDT involves the exposure of a photosensitiser (PS) to a specific light source in the presence of molecular oxygen, so that reactive oxygen species (ROS) are generated that induce the killing of pathogens. Despite their short lifetime, ROS can quickly react with multiple structural components of bacteria (lipids, proteins, nucleic acids), causing irreversible and diverse damage, which is key to limiting the rise of APDT-resistant bacteria [12,13]. This makes ADPT far less susceptible to resistance-based complications than traditional antibiotics. Systematic variation of PS dosage, light intensity and exposure time also provides an experimental space to control the extent of the photodynamic effect towards bacterial inactivation and preserved cell viability.

Despite the aforementioned features, the translation of ADPT to the clinic remains challenging. The PS is typically delivered to the infected site in an aqueous solution state, triggering inherent issues of PS-induced tissue staining and aesthetic discomfort for the patient, as well as risks of suboptimal photodynamic response due to concentration-dependent aggregation of PS molecules [14]. Given their relatively low molecular weight, PS molecules can also quickly diffuse out of the local target, limiting the localised generation of ROS, potentially causing side effects and further tissue staining [15].

The confinement of clinically approved PSs in polymer carriers offers a regulatory-friendly, scalable and quicker route to minimise risks of PS diffusion, aggregation and tissue staining, compared to the synthesis of new PSs [16,17]. This approach also enables the possibility to equip medical device prototypes with photodynamic capabilities, aiming to accomplish enhanced and selective antimicrobial effects together with specific end-use functionalities [18,19]. Strategies enabling confinement of soluble factors have been reported via polymer solution spinning, e.g. electrospinning [20] and wet spinning [21], solution casting [22] and secondary interactions [23,24,25]. These studies have shown that the release of the PS can be adjusted by variation in microstructure and chemical composition of the polymer carrier, or through molecular interactions between the soluble factor and the polymer carrier [26]. On the other hand, a relatively high dosage of the PS is still required to offset the release of the PS over time, which could potentially result in unwanted alteration of the wound environment, toxicity and short-lived photodynamic capability. Other confinement strategies have looked at the conjugation of PSs to a substrate, whereby the PS molecule has been successfully grafted to synthetic [27,28] and natural [29,30,31,32] polymers. The covalent functionalisation of the PS raises questions on the controllability of the photodynamic behaviour of the respective PS-conjugated system, so that significant research has been devoted to the chemical characterisation and preclinical testing of resulting products. On the other hand, minimal attention has been paid to the processability of the resulting PS-conjugated systems for integration into wound dressings, which is key to foster technology translation towards clinical impact.

The aim of this work was to accomplish a novel PVA derivative with photodynamic and photocurable functionalities aiming to integrate antibiotic- and staining-free infection control into relevant wound dressing formats while ensuring cell viability. Leveraging our expertise in biopolymer-based molecular networks [33,34,35,36] and textile fibre manufacturing [37,38,39,40], we hypothesised that the conjugation of PVA with both phenothiazine and methacrylate residues would enable long-lasting PS retention and photodynamic capability against chronic wound bacteria, together with mammalian cell tolerability and broad polymer processability. In the following, a molecular design approach of significant novelty and impact is presented, in which PVA [41,42] is selected as a chemically accessible and clinically approved building block [43,44,45] for the creation of a high-value photodynamic material with broad healthcare applicability, from chronic wound dressings up to minimally invasive localised therapy.

## 2. Materials and Methods

### 2.1 Chemicals

PVA ($M_w$: 146,000–186,000 g mol$^{-1}$, 99% hydrolysed), p-Toluenesulfonyl chloride (TsCl), TB, methacrylic anhydride (MA), triethylamine (TEA) and 2-Hydroxy-4'-(2-hydroxyethoxy)-2-methylpropiophenone (I2959) were purchased from Sigma Aldrich. Acetone, ethanol and 1,1,1,3,3,3-Hexafluoro-2-propanol (HFIP) were purchased from VWR, Sigma Aldrich, and Fluorochem, respectively. Minimum Essential Medium Eagle, Alpha Modification (alpha-MEM), trypsin, foetal bovine serum (FBS), L-glutamine (Gln), and penicillin-streptomycin (P-S) were purchased from Sigma Aldrich. Phosphate buffered saline (PBS) was purchased from Lonza (Slough, UK). Calcein-AM/ethidium homodimer Live/Dead assay and alamarBlue™ Cell Viability Reagent were purchased from Thermo Fisher.

### 2.2 Synthesis of functionalised PVA products

*Tosylated PVA (PVA-Ts)*. PVA was dissolved in distilled water (3 wt.% PVA) at 90 °C under magnetic stirring. The solution was cooled down to room temperature under stirring, so that TsCl was added at a molar ratio of 50 mol.% with respect to the vinyl alcohol repeat unit. Following the introduction of TEA (1 eq. of hydroxyl groups), the solution temperature was increased to 90 °C and the reaction ran for 24 hours under magnetic stirring. The reaction mixture was subsequently cooled down to room temperature and precipitated in a 10-fold excess of acetone. Following overnight incubation, the pellet was recovered via centrifugation (10000 xg, 10 min) and dried at 60 °C.

*TB-conjugated PVA (PVA-TB).* PVA-Ts was dissolved in distilled water (3 wt.% PVA-Ts) under magnetic stirring at 90 °C. The solution was cooled down to room temperature under stirring, prior to addition of TB (0.5 mol.% of vinyl alcohol repeat unit) and TEA (1 eq. of hydroxyl groups). The temperature was subsequently increased to 90 °C and the reaction ran for 48 hours under magnetic stirring. The reaction mixture was subsequently cooled down to room temperature and precipitated in a 10-fold excess of acetone. The product was recovered by centrifugation, dissolved again in distilled water (at 90 °C) and re-precipitated in acetone. This purification procedure was conducted three times to remove any unreacted dye. The purified pellet was collected by centrifugation and dried at 60 ºC.

*TB-conjugated PVA methacrylate (PVA-TB-MA)*. PVA-TB was dissolved in distilled water (3 wt.% PVA-TB) at 60 °C under magnetic stirring, prior to addition of MA (50 mol.% of vinyl alcohol repeat unit) and TEA (1 eq. of hydroxyl groups). The mixture was reacted at 60 °C for 5 hours under magnetic stirring and then precipitated in a 10-fold excess of acetone. The product was purified and recovered as reported above, followed by drying at 60 °C.

*PVA methacrylate (PVA-MA)*. PVA was dissolved in distilled water (3 wt.% polymer) at 90 °C under magnetic stirring, prior to addition of MA (50 mol.% of vinyl alcohol repeat unit) and TEA (1 eq. of hydroxyl groups). The mixture was reacted at 60 °C for 5 hours under magnetic stirring and then precipitated in a 10-fold excess of acetone. The product was purified and recovered as previously reported, prior to drying at 60 °C.

**2.3 Chemical characterisation**

Attenuated total reflection Fourier transform infrared (ATR–FTIR) spectra were recorded with dry polymer samples using a Spectrum One FT-IR Spectrometer (PerkinElmer, Waltham, MA, USA) with a Golden Gate ATR attachment (Specac Ltd., London, UK). Scans were conducted from 4000 to 600 $cm^{-1}$ with 100 repetitions averaged for each spectrum. Resolution and scanning intervals were 4 $cm^{-1}$ and 2 $cm^{-1}$, respectively.

Proton nuclear magnetic resonance ($^1$H-NMR) spectra were recorded (500 MHz, Bruker) by dissolving 16 mg of polymer in 2 mL of deuterium oxide ($D_2O$, Sigma Aldrich) at 90 °C in a small glass vial. The solution was cooled down to room temperature, prior to recording the $^1$H-NMR spectra at room temperature.

Absorption spectra of native PVA and functionalised products PVA-TB and PVA-TB-MA were measured using a microplate reader (Varioskan LUX, Thermo Scientific). Full spectra were recorded for the polymer solutions (10 $mg \cdot mL^{-1}$) and the TB-supplemented solution controls (n=2). The absorbance values of the control solutions of PVA in distilled water (n=2) were subtracted from the absorbance values measured with the aqueous solutions of PVA-TB and PVA-TB-MA. Likewise, the absorbance values of distilled water were subtracted from the absorbance values of the TB-supplemented solutions in distilled water. Linear fitting ($R^2$> 0.99) of the absorbance values revealed by the TB-supplemented solutions at either 590 nm or 615 nm generated two calibration curves (Figure S1, Supp. Inf.), which were used for the quantification of TB coupling in PVA-TB and PVA-TB-MA, respectively.

### 2.4 Preparation of PVA-based films

PVA-TB-MA was solubilised (3 wt.%) in HFIP via magnetic stirring at room temperature in the dark. The polymer solution (either 0.5 or 0.2 mL) was cast into silicone moulds of varied size (Ø: 15 mm, h: 3 mm; or Ø: 11 mm, h: 3 mm) and left to air dry in a chemical fume hood for up to 48 hours at room temperature prior to further use.

### 2.5 Wet spinning of individual polymer fibres

The wet spinning solutions of PVA (10-15 wt.%) and PVA-TB (12 wt.%) were prepared in distilled water by magnetic stirring at 90 °C. The wet spinning solutions of PVA-TB-MA (12 wt.%) were prepared in HFIP by magnetic stirring at room temperature. The resulting solutions were equilibrated to room temperature, prior to being transferred to a syringe equipped with a needle (0.8 mm internal diameter). Once that the syringe was loaded on to a syringe pump (AL-1000, WPI), wet spinning was carried out at varying flow rates (2–8 $mL\cdot hr^{-1}$) using acetone as the coagulation bath. The wet spun fibres were kept submerged in a sealed coagulation bath for 1 hour, prior to fibre collection and air-drying. Fibres made of native PVA were labelled as *PVAX*, where *X* denotes the polymer concentration in the wet spinning solution. Fibres made of PVA-TB and PVA-TB-MA were labelled as F-TB and F-TB-MA, respectively.

**2.6 UV-curing of polymer samples**

I2959 was dissolved (1 wt.%) my magnetic stirring in the dark for 30 min either in acetone at room temperature or in distilled water at 90 °C. Following solution equilibration to room temperature, samples of cast film and wet spun fibre were incubated in the I2959-supplemented solution and UV-cured (365 nm, Chromato-Vue C-71, Analytik Jena, Upland, CA, USA). Films were irradiated for 30 min on the top and the bottom side, while fibres were irradiated for 30 min on the top side, only. UV-cured samples were collected and washed with acetone (twice), prior to 24-hour air-drying. Films cured in either acetone or distilled water were labelled as A-TB-MA* and H-TB-MA*, respectively. Fibres cured in either acetone or distilled water were labelled as FA-TB-MA* and FH-TB-MA*, respectively.

**2.7 Preparation of UV-cured dressing composite**

Samples of PVA-TB-MA were dissolved (5 wt.% polymer) in an I2959-supplemented solution of distilled water (1 wt.% I2959) under magnetic stirring at 80 °C, and equilibrated to room temperature prior to integration with a collagen-based dressing prototype [46]. The dressing precursor (labelled as 4VBC) was prepared via the reaction of type I bovine atelocollagen (Collagen Solutions PLC) with 4-vinylbenzyl chloride. The resulting photoactive product revealed a degree of functionalisation of 22±2 mol.% (n=2), in line with previous reports [47,48]. 4VBC was dissolved (0.8 wt.%) in a 10 mM hydrochloric acid solution supplemented with 1 wt.% I2959, under magnetic stirring at room temperature. The solution of 4VBC was cast on a 96-well plate (0.08 g per well), followed by the addition of the PVA-TB-MA solution (0.06 g). The top and bottom sides of the sample were subsequently exposed to UV light (365 nm, Chromato-Vue C-71, Analytik Jena, Upland, CA, USA) for 6 min at each side, to generate a two-layer composite, labelled as C-PVA*. Control samples of either UV-cured PVA-TB-MA or UV-cured 4VBC were also prepared and labelled as PVA-TB-MA* and 4VBC*, respectively. All samples were thoroughly washed in distilled water prior to further use.

**2.8 Mechanical tests**

Rheological measurements were carried out on water-swollen UV-cured films (Ø: 30 mm, h: 1.5 mm) using an MCR 301 rheometer (Anton Paar, Graz, Austria). An amplitude sweep was conducted to identify the linear-elastic region using an angular frequency of 10 $rad \cdot s^{-1}$, followed by frequency sweeps at room temperature using a constant amplitude of 1 %.

Tensile tests (INSTRON 5544, 10 N loading cell) were carried out on individual dry fibres (n=5, length: 5 mm) using an elongation rate of 20 $mm \cdot s^{-1}$ and a gauge length of 10 mm. The resulting stress-strain curves were fitted linearly (2–5% strain) to quantify the Young's modulus. Results were reported as mean ± standard deviation.

Freshly synthesised water-equilibrated discs (Ø = 7 mm, $n$ = 8) were compressed at room temperature with a compression rate of 3 $mm \cdot min^{-1}$ (BOSE EnduraTEC ELF 3200, EnduraTEC Systems Corporation, Minnetonka, MN, USA) using a 10 N load cell. The compression modulus was quantified by linear fitting in the compression range of 20-30%.

**2.9 Gel content and swelling tests**

The gel content ($G$) was measured to confirm UV-induced network formation and quantify the overall portion of water-insoluble sample. Dry UV-cured samples ($m_d$: 5–7 mg, n=3) were individually incubated in 1 mL of distilled water (40 °C, 24 hours), prior to air-drying and weighing ($m_1$). $G$ was calculated through equation 1:

$$G = \frac{m_1}{m_d} \times 100 \quad \textbf{(Eq. 1)}$$

Swelling tests were performed on dry UV-cured fibres via incubation in 1 mL of either deionised water or PBS (10 mM, pH 7.4, 37 °C), followed by gravimetric and dimensional analysis. The weights of dry samples ($m_d$) and water-equilibrated samples ($m_s$) were recorded and the swelling ratio (*SR*, n=3) calculated through the following equation:

$$SR = \frac{m_s - m_d}{m_d} \times 100 \qquad \textbf{(Eq. 2)}$$

The swelling index (*SI*) was quantified by measuring the fibre diameters in the dry state and following 24-hour incubation in either deionised water or PBS (10 mM, pH 7.4, 37 °C). Optical microscopy images (OLYMPUS BX60) were captured prior to the measurements of fibre diameter across five different locations over a 1.5–2 cm fibre length. The *SI* was calculated through the following equation:

$$SI = \frac{\emptyset_s}{\emptyset_d} \times 100 \qquad \textbf{(Eq. 3)}$$

where $\emptyset_s$ and $\emptyset_d$ indicate the fibre diameters in the swollen and wet state, respectively.

**2.10 PS release tests**

Dry UV-cured films of A-TB-MA* (15±2 mg, n=3) and TB-loaded controls, i.e. PVA-MA*(TB2) (10±1 mg, n=3) and PVA-MA*(TB10) (11±1 mg, n=3), were individually incubated at room temperature in 3 mL of PBS (10 mM, pH 7.4, 25 °C). Controls were prepared by loading a solution of PVA-MA (3 wt.% in HFIP) with TB, followed by UV-curing in acetone and air drying (see section 2.6). PVA-MA*(TB2) and PVA-MA*(TB10) contained 2 wt.% and 10 wt.% of TB, respectively. The amount of TB released in PBS was determined via UV-Vis spectrophotometry over 96 hours. The absorbance values were subtracted from the absorbance value of the sample-free PBS solution. A calibration curve was generated in PBS by recording the absorbance at 610 nm (Figure S2, Supp. Inf). The absorbance values were subtracted from the absorbance values of the TB-free PBS solution.

**2.11 ROS fluorometric assay**

20 mM stock solution was prepared by adding 9.74 mg of 2',7'–dichlorodihydrofluorescein diacetate (DCFDA, Invitrogen) in 1 mL of absolute ethanol. 2mM of working solution was prepared by adding 0.1 mL stock solution to 0.9 mL of absolute ethanol and was covered with aluminium foil to avoid contact with light. TB and PBS were used as positive and negative controls (n=3), respectively. UV-cured films A-TB-MA* (n=3) of known dry weight ($m_d$: 5-7 mg) were incubated in a 24-well plate containing 2 mL of PBS in each well and subjected to work light exposure (6000-lumen, 50W, 135 lumen/W, 2800-3200 warm light) for either 30 or 60 min. At each time point, 200 µL of the light-irritated solution and 2 µL of the aforementioned 2 mM DCFH-DA working solution were added to a 96-well dark plate and stored in the dark for one hour. The intensity of fluorescence, corresponding to the generation of ROS, was quantified using a Varioskan LUX MulKmode Microplate Reader with 485 nm and 530nm wavelengths for excitation and emission, respectively.

## 2.12 Antibacterial tests *in vitro*

*Bacterial culture.* Bacterial cultures were prepared overnight via aseptic addition of 5 mL of Mueller Hinton Broth (MH) media to a sterile universal tube. A sterilised culture loop was used to collect a single bacterial colony from an agar plate containing either *S. aureus* NCTC 8532 or *P. aeruginosa* NCTC 10332 bacterial strains. Each bacterial colony was added to the test tube, prior to overnight aerobic incubation at 37°C. An absorbance reading of 0.08 was considered to equate to $1.5\times10^{8}$ $CFU\cdot mL^{-1}$ according to the McFarland 0.5 standard used to calculate the number of bacterial cells in the overnight culture solution.

*Sample preparation*. UV-cured samples (Ø ~10 mm, $m_d$ ~15 mg) were disinfected for 15 minutes on each side via UV light exposure. Samples (n=3) were subsequently hydrated (30 min) in a 24-well plate containing 1 mL of sterile PBS in each well, prior to transfer to 1 mL of $1\times10^{6}$ $CFU\cdot mL^{-1}$ suspension of either *S. aureus* NCTC 8532 or *P. aeruginosa* NCTC 10332.

*Light activation.* Well plates containing the samples were irradiated with a work light (6000-lumen, 50W, 135 lumen/W, 2800-3200 warm light) for either 30 or 60 min. Well plates were subsequently incubated aerobically for four hours at 37 ˚C. Remaining live bacteria ($CFU\cdot mL^{-1}$) were calculated through the dilution plating method and overnight incubation.

*Scanning Electron Microscopy (SEM)*. Tested samples were gently washed with sterile PBS, fixed in 4% glutaraldehyde overnight, and dehydrated in distilled water solutions of increasing concentration of ethanol (30, 50, 70, 90, 95, 100 vol.% EtOH). Bacterial morphology was imaged with a Hitachi S-3400N microscope (Hitachi, Tokyo, Japan).

### 2.13 Cytotoxicity study *in vitro*

*Cell culture*. L929 mouse fibroblasts (passage number 15) were cultured (37°C, 5% $CO_2$) in a T-150 cell culture flask with alpha-MEM supplemented with 10% FBS, 1% P-S, and 1% Gln.

*Sample and extract preparation*. UV-cured wet spun fibres FA-TB-MA* and UV-cured films A-TB-MA* were incubated in 70 vol.% ethanol and subsequently air-dried in the dark prior to use. Extracts were prepared by incubation of 75 mg of ethanol-treated samples in 2 mL of alpha-MEM, prior to 60 min irradiation with a 6000-lumen work light (850 W, 6000-6500 K) located at a 57 mm distance from the vial. Samples were incubated for 24 hours at 37°C.

*Contact cytotoxicity tests.* Ethanol-treated samples were individually transferred to the wells of a 96-well plate, prior to L929 cell seeding ($1 \times 10^4$ cells per well). The samples were then irradiated with a 6000-lumen work light (850 W, 6000-6500 K), as described above.

*Extract cytotoxicity tests*. L929 fibroblasts were seeded in a 96-well plate containing 100 µL alpha-MEM ($1 \times 10^4$ cells per well). 40 µL of extract were added to each of the test wells, prior to 24-hour incubation at 37 °C.

*Cellular metabolic activity*. Following 24- and 72-hour cell culture, media were removed from each well and replaced with fresh media supplemented with 10% of AlamarBlue reagent. The cells were then incubated for a further 2 hours at 37 °C. Fluorescence was measured using a Thermo Scientific Varioskan Lux plate reader with excitation at 560 nm and emission at 590 nm. Fluorescence readings were analysed using OriginPro 2023b software. Each sample was tested in triplicate. Cellular tolerability was assessed based on a viability threshold of 70% (ISO 10993-5:2019).

*Live/dead staining*. Following 24- and 72-hour cell culture, media were removed from each well and the wells washed twice with PBS. A 50 µL solution of live/dead stain was added to each well, whereby calcein-AM and ethidium homodimer were diluted to 2 and 4 µM, respectively, before use. The cells were incubated for 45 minutes at 37 °C. After incubation, the wells were washed twice with PBS. Live/dead images were captured using a Leica TCS

SP8 confocal microscope at 10x magnification.

### 2.14. Statistical analysis

Statistical analysis was carried out using Origin software (version 2024b Academic). Statistical significance was determined with one-way ANOVA corrected with Bonferroni test. Data were presented as mean ± standard deviation.

## 3. Results and discussion

Figure 1 describes the synthetic approach employed to accomplish the PVA derivative equipped with staining-free photodynamic capability and broad processability. Unlike other phenothiazines used in APDT, e.g. methylene blue [49], the presence of the free primary amine in TB offers a relatively simple route for chemical functionalisation, including conjugation to proteins [50], glycopolymers [51], and biodegradable polymers, such as PVA.

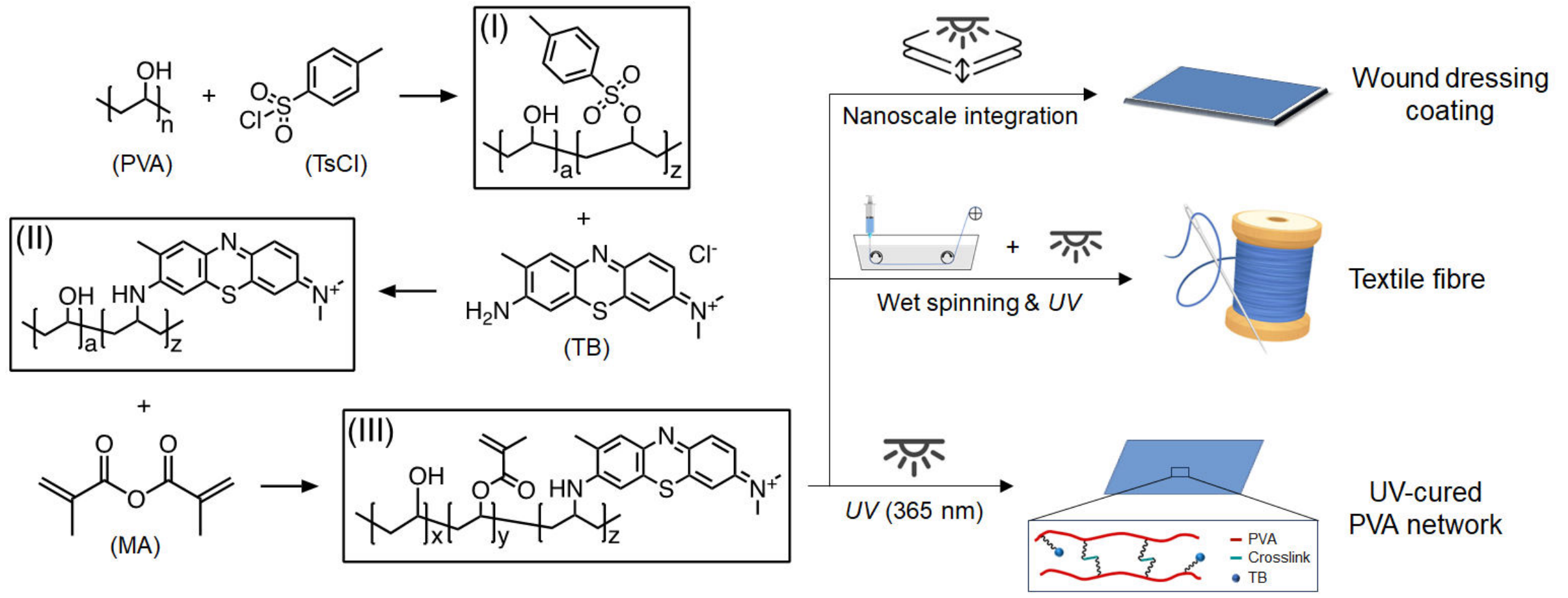

**Figure 1.** Synthesis of staining-free photodynamic PVA and UV-cured network with bespoke material format. Tosylation of PVA (I) enables the grafting of TB residues (II). The dyed product (PVA-TB) is subsequently reacted with MA to accomplish a PVA derivative bearing both phenothiazine and methacrylate residues (PVA-TB-MA, III). The resulting UV-cured product can be delivered as a cast film, a wet spun textile fibre, and as a nanoscale coating for the design of antimicrobial wound dressings.

To realise this, tosylation of PVA was selected (I), given the reactivity of TsCl towards the hydroxyl groups of PVA, the stability of tosylated PVA (PVA-OTs) under ambient conditions, and the tosylate leaving group capability to allow further PVA modification [52]. Subsequently, a nucleophilic substitution reaction between PVA-OTs and TB was pursued (II), exploiting the reactivity of the primary amino group in the phenothiazine residue. Reaction of the resulting product (PVA-TB) with MA completed the synthesis, aiming to generate TB-conjugated PVA methacrylate (PVA-TB-MA) with UV-curable residues on the polymer backbone (III). In this way, the molar content of the photodynamic residues of TB was expected to be independently regulated from the molar content of the methacrylate moieties of MA, given the impact of these on the antimicrobial photodynamic effect, and photocuring capability, respectively.

**3.1 Chemical characterisation of PVA derivatives**

FTIR spectroscopy was performed to analyse the chemical composition of native samples and reacted products. The characteristic absorption bands of PVA were detected in the spectrum of the native polymer (Figure 2A), whereby the broad signal at 3270 $cm^{-1}$ was attributed to the stretching vibrations of the O–H groups, while the signal at 2920 $cm^{-1}$ supported the presence of C–H bonds. The absorption bands of $CH_2$ groups were detected at 1420 $cm^{-1}$ and 1330 $cm^{-1}$ together with strong C-O stretching vibrations at 1087 $cm^{-1}$, associated with C-O-H groups [53]. The signal at 1142 $cm^{-1}$ was attributed to crystalline C=O stretching due to the semi-crystalline nature of PVA [54,55].

Tosylation of PVA led to the detection of additional bands in the FTIR spectrum of sample PVA-OTs (Figure 2B), i.e. at 1398 and 1175 $cm^{-1}$, corresponding to the vibrations of S=O bonds, and at 820 $cm^{-1}$, related to the presence of C–O–S bonds [56]. New signals were also found at 680 $cm^{-1}$, which was attributed to the out-of-plane C–H bending vibrations, and at 1650 $cm^{-1}$, which was assigned to the aromatic C=C stretching groups.

Further to the reaction with TB, the FTIR spectrum of sample PVA-TB (Figure 2C) revealed the disappearance of the characteristic tosyl-related bands at 820 $cm^{-1}$ and 680 $cm^{-1}$, providing indirect evidence of the nucleophilic substitution of Ts residues with TB (Figure 1). The band at 1650 $cm^{-1}$ was attributed to the vibrations of the C=C aromatic bonds of TB [57,58,59], as observed in the FTIR spectrum of respective raw material (Figure 2E). On the other hand, the crystallisation-sensitive FTIR band of native PVA at 1142 $cm^{-1}$ was not detected with sample PVA-TB (Table S1, Supp. Inf.). This indicates a significant reduction in polymer crystallinity, in line with the introduction of the bulky TB residues on the PVA backbone.

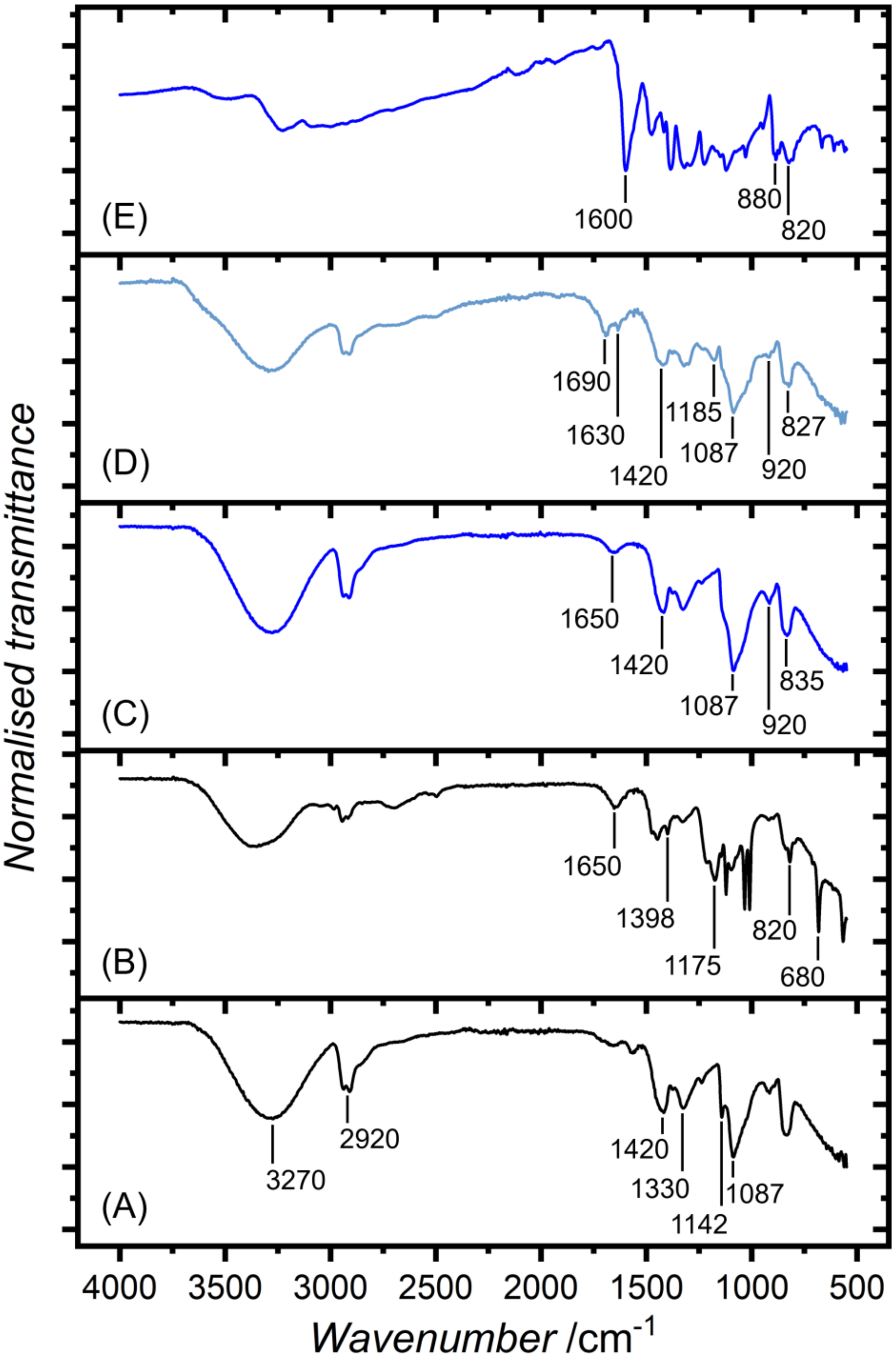

**Figure 2.** FTIR spectra of raw materials and functionalised products. (A): PVA; (B): PVA-OTs, (C): PVA-TB, (D): PVA-TB-MA, (E): TB.

Following reaction with MA, the FTIR signal associated with the C=C aromatic bonds of the phenotiazine residues shifted from 1650 to 1630 cm$^{-1}$. Furthermore, two new signals were observed at 1690 cm$^{-1}$ and 1185 cm$^{-1}$ in the FTIR spectrum of PVA-TB-MA (Figure 2D), describing the C=C and ester OC–O–C stretching vibrations, respectively, in line with the grafting of the methacrylate residues to the PVA. The coupling of MA residues to the free hydroxyl groups of PVA-TB (Figure 1) was also indirectly supported by the presence of a weaker band at 3270 $cm^{-1}$ and the absence of the crystallisation-sensitive band at 1142 $cm^{-1}$ (Figure 2 and Table S1, Supp. Inf.).

To achieve direct confirmation of the coupling of both TB and MA onto the polymer backbone, $^1$H-NMR spectra of the reacted products were recorded. The spectra of native PVA revealed the characteristic peaks at ~1.6 ppm and 3.9 ppm (Figure S3, Supp. Inf.), corresponding to the hydrogens of $CH_2$ and CH residues, respectively [40,41]. Following reaction with TsCl, additional peaks were observed at 7.2-7.6 ppm and 2.3 ppm in the spectrum of PVA-OTs (Figure 3A), consistently with the presence of aromatic and methyl protons in grafted tosylate residues, respectively [60]. Peak integration of the tosylate-related signals to either proton signals of PVA indicated a degree of tosylation of up to 17 mol.% (Table S2, Supp. Inf.). Interestingly, higher values of tosylation were measured from the methyl, rather than aromatic, signal of the tosylate residue (Figure S3, Supp. Inf.). The most likely explanation of this observation is the overlap of the methyl signal (at 2.3 ppm) from the tosylate residue with the acetate signal (at ~2 ppm) from PVA.

Further to the reaction of PVA-OTs with TB, the resulting product revealed new $^1$H-NMR signals at 7.2-7.7 ppm and 2.1 ppm (Figure 3B, zoomed-in regions), which were associated with the aromatic and methyl residues of TB (Figure 3D), respectively [50,61]. Despite the above evidence of TB grafting, the relatively low intensity of the phenothiazine aromatic signals in the $^1$H-NMR spectrum of PVA-TB did not agree with the corresponding suppression of the tosylate signals (Figure 1). This observation was attributed to the water-induced product detosylation [62], consequent to the relatively high reaction temperature and relatively long reaction time (90 °C, 24 hours). This hypothesis was supported by control reactions performed with a reduced molar ratio of both TsCl ($[Ts]\cdot[OH]^{-1}= 0.1$) and TB ($[TB]\cdot[Ts]^{-1}= 0.05$), which confirmed the successful tosylation and detosylation of PVA, together with the presence of weak phenothiazine aromatic signals (Figure S4A and Table S2, Supp. Inf.).

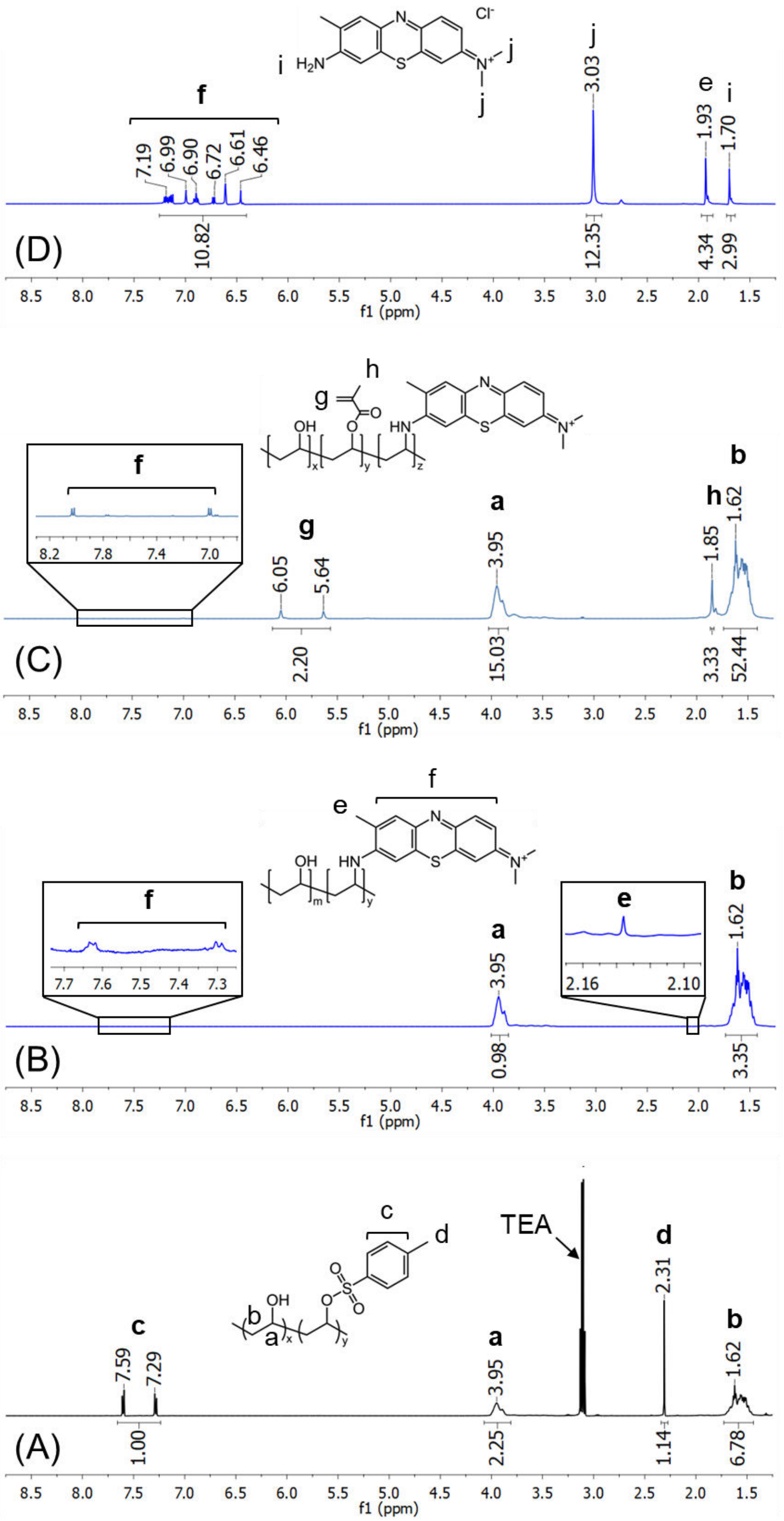

**Figure 3.** $^1$H-NMR spectra (500 MHz, $D_2O$, 25 °C) of PVA-OTs (A), PVA-TB (B), PVA-TB-MA (C) and TB (D).

When sample PVA-TB was subsequently reacted with MA, the $^1$H-NMR spectrum of the product revealed new peaks at 5.6-6.1 ppm and at 1.9 ppm (Figure 3C), corresponding to the vinyl and methyl protons of the methacrylate residues, respectively [33,63,64]. Integration of either MA-related signal with respect to either proton signals of PVA revealed a degree of methacrylation of up to 7 mol.%, which was consistent with the value measured in the control sample (Figure S4C and Table S2, Supp. Inf.) and in previously reported PVA derivatives [65]. The consistent molar content of MA residues in samples of PVA-TB-MA synthesised from precursors with varied degrees of tosylation (Table S2, Supp. Inf.) supports the reproducibility of this synthetic route and indicates the possibility to control the molar content of methacrylate residues independently of the chemical composition of the PVA-TB precursor.

Other than $^1$H-NMR, products PVA-TB and PVA-TB-MA were also characterised by UV-Vis spectroscopy (Figure 4A), aiming to obtain quantitative confirmation of TB conjugation.

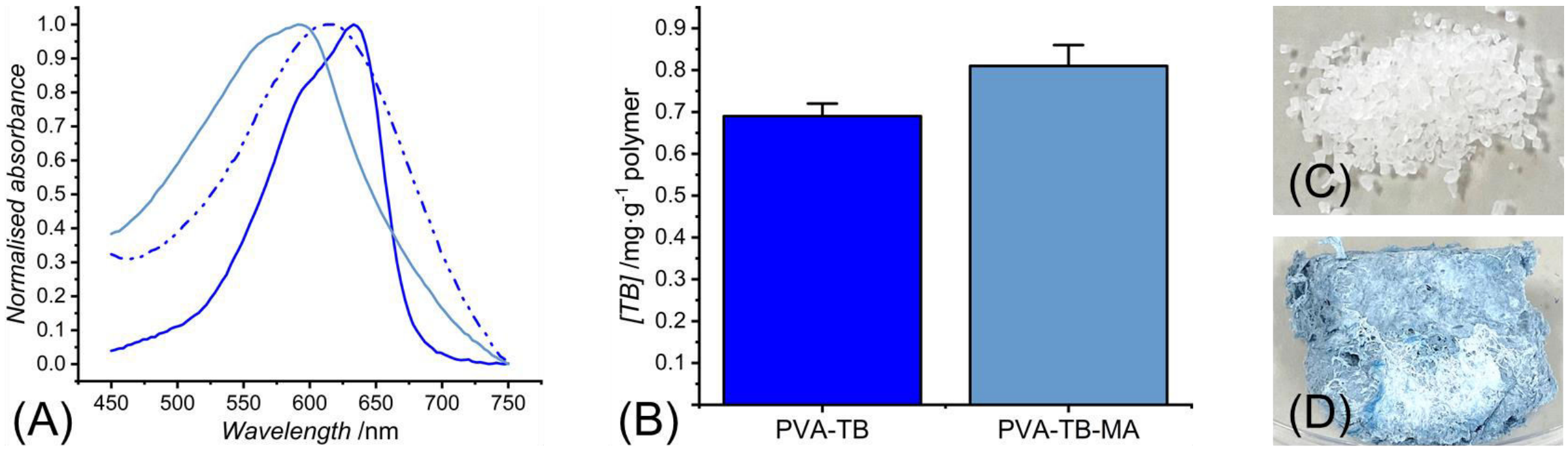

**Figure 4.** Quantification of TB grafting in, and appearance of, TB-reacted samples. (A): UV-Vis spectra of distilled water (1 mL) supplemented with either TB (0.031 mg, blue solid), PVA-TB (10 mg, blue dash dot dot) or PVA-TB-MA (10 mg, light blue solid). Data represent the mean value of the normalised absorbance (n=2). (B): Calculated weight fraction of TB conjugated to PVA products. Data are presented as mean ± standard deviation (n=2). (C-D): Photographs of samples of native PVA (B) and PVA-TB (C).

The absorption maximum of TB-supplemented solution controls was measured at 633 nm, in line with previous reports [61]. On the other hand, the UV-Vis spectra of PVA-TB and PVA-TB-MA revealed a blue shift in the absorption maxima, i.e. from 633 nm to 615 nm and 591 nm, respectively, which agrees with the binding of the phenothiazine residues to the polymer [50,66]. UV-Vis calibration curves built with varied photosensitiser concentrations from the absorbance readings at 615 nm and 591 nm (Figure S1, Supp. Inf.) enabled the quantification of the weight fraction of TB conjugated to both PVA-TB and PVA-TB-MA (Figure 4B). This equated to 0.69±0.03 and 0.81±0.05 mg of TB per gram of polymer, respectively, whereby the insignificant variation recorded in the two samples supports the binding, rather than the physical absorption, of TB to the PVA backbone. The information revealed by the UV-Vis spectra was consistent with the significant colour difference observed with the samples of PVA (white, Figure 4C) compared to PVA-TB (blue, Figure 4D), again supporting the coupling of phenothiazine residues to the polymer.

### 3.2 Physical properties of UV-cured networks

Following confirmation of the chemical structure of PVA-TB-MA, the attention moved to the synthesis of the corresponding UV-cured network and corresponding cast films. Extraction in distilled water (40 °C, 24 hours) was initially carried out to assess the release of any non-crosslinked polymer fragments (Figure 5A). A high gel content (*G*) was measured in samples cured in either acetone (*G*= 96±12 wt.%) or distilled water (*G*= 99±6 wt.%), in line with the complete gelation observed when the photoactive polymer was dissolved in water prior to UV exposure, and the formation of covalent crosslinks at the molecular scale. Frequency sweeps on the water-equilibrated networks were subsequently carried out (Figure 5B), revealing a predominantly elastic behaviour with higher storage modulus (*G'*: 8-12 kPa) compared to the loss modulus (*G''*: 1-5 kPa). These results are comparable to previously reported PVA-based hydrogels [43,44] and further confirm the synthesis of a covalently crosslinked polymer network following UV exposure [33,67].

On the macroscopic scale, the UV-cured networks revealed a blue colour even following equilibration in water (Figure 5B, inset), supporting the fact that the PS remained confined within, and did not leach away from, the hydrogel structure, in agreement with the binding of the TB to the PVA. To confirm TB confinement within the polymer network, samples of A-TB-MA* were incubated in PBS alongside controls PVA-MA*(TB2) and PVA-MA*(TB10), containing 2 wt.% and 10 wt.% of dispersed TB, respectively. After 96-hour incubation in PBS, the UV-cured TB-conjugated sample indicated minimal release of TB (3.04±0.42 wt.%), equating to an overall mean weight value of 0.37 µg (Figure 5C).

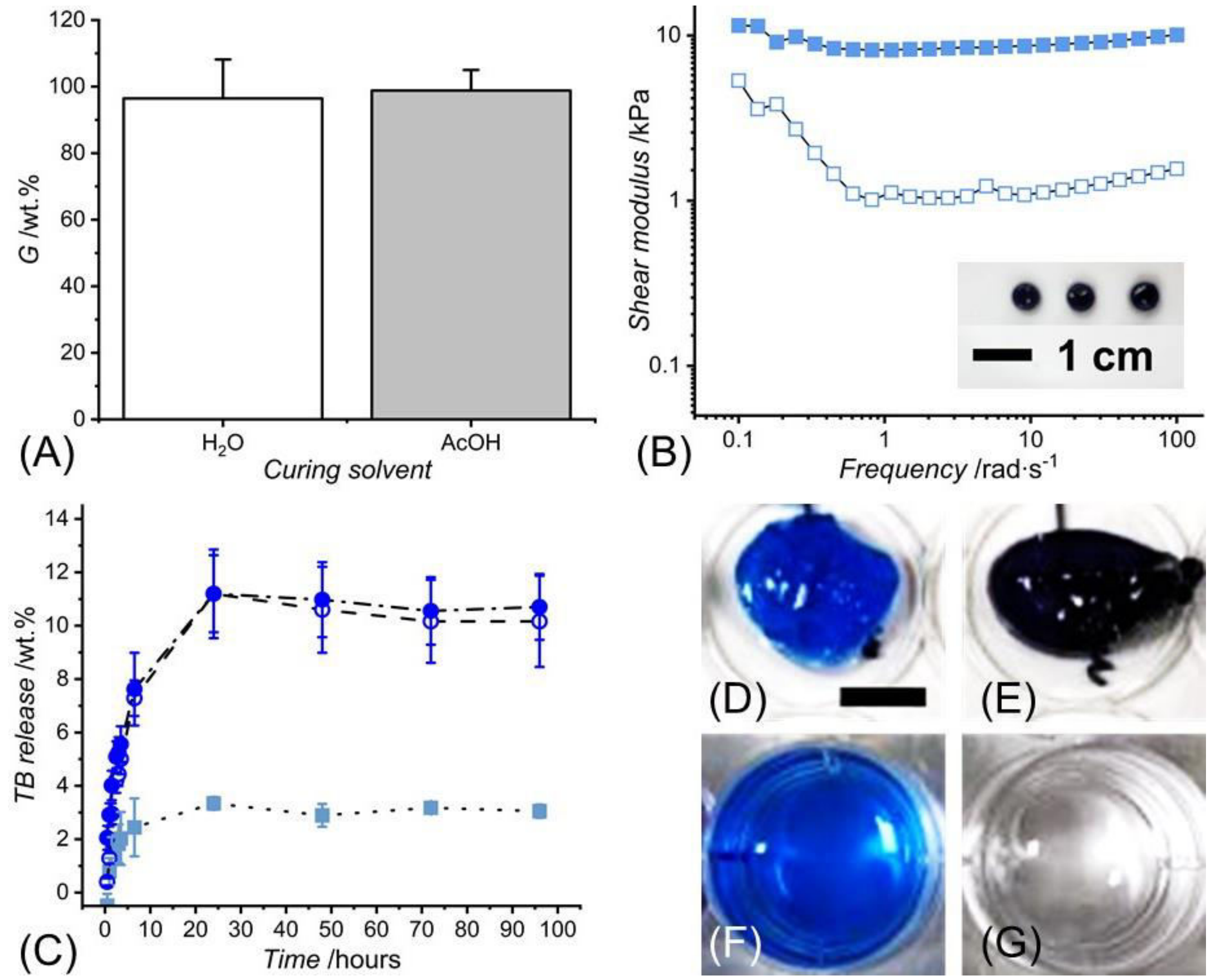

**Figure 5.** UV-cured staining-free network formation. (A): gel content (*G*) of films of PVA-TB-MA cured in distilled water ($H_2O$, white) and acetone (AcOH, grey). Data are presented as mean ± standard deviation (n=3). (B): Typical frequency sweep of the water-equilibrated acetone-cured sample A-TB-MA*. (■): Storage modulus (*G'*); (□): loss modulus (*G''*). The inset depicts the photograph of three replicates. (C): TB release measured during a 96-hour incubation (10 mM PBS, pH 7.4, 25 °C) of sample A-TB-MA* (■) and TB-loaded controls PVA-MA*(TB2) (○) and PVA-MA*(TB10) (●). Data are presented as mean ± standard deviation (n=3). (D-G): Photographs of PBS-incubated samples (96 h, 25 °C) of PVA-MA*(TB2) (D) and A-TB-MA* (E), as well as respective supernatants (F-G). The scale bar (≈ 0.5 cm) applies to all photos.

In contrast, a mean release of about 11 wt.% was observed after 96-hour incubation of the UV-cured TB-loaded controls, equating to ca. 20 µg and 107 µg from PVA-MA*(TB2) and PVA-MA*(TB10), respectively. These results agree with the colour fading of the material (Figure 5D-E) and the significant colouration of the supernatant (Figure 5F-G) detected after the 96-hour incubation of the TB-loaded controls, in contrast to the case of the TB-conjugated sample. These observations, together with the suppressed release of TB measured with sample A-TB-MA* compared to the TB-loaded controls, confirm the suppression of TB leaching *in vitro* and the successful conjugation of TB to PVA.

### 3.3 Fibre morphology and spectroscopic analysis of wet spun products

Having demonstrated the photocuring capability of, and dye retention in, the TB-conjugated PVA derivatives, the polymer processability into textile fibres was investigated, given the applicability of these in wound dressings and other healthcare products [37-40]. Wet spinning was selected as an industry-compliant, scalable manufacturing process, which yields individual and customisable fibres, and which does not require high temperatures for polymer extrusion [68]. Due to the high melting temperature of PVA, wet spinning was considered a more eco-friendly alternative to melt spinning [69]. The fact that the solubility of PVA is limited to aqueous and highly polar solvents is also appealing, aiming to accomplish fibre coagulation in relatively benign organic solvents, such as acetone.

Relatively small variation in PVA concentration (10–12 wt.%) in the wet spinning solution proved to generate homogeneous fibres with insignificantly different diameters (Ø: 282±29–281±41 μm) (Figure S5, Supp. Inf.). A further increase in PVA concentration yielded detectable irregularities in fibre morphology as well as significantly increased fibre diameters (Ø= 493±84 μm). The concentration of the polymer in the wet spinning dope is known to have a direct effect on the molecular interactions and adhesion between polymer chains, due to e.g. hydrogen bonding [70,71], which are key to promoting the coagulation of wet spun fibres. However, upon further increase in the concentration of PVA, the molecular interactions between fibre-forming polymer chains become too great, generating significant shear forces and flow resistance. These phenomena act against the molecular alignment of polymer chains during wet spinning, yielding dense and large-diameter fibres with irregular morphologies [37,38,72].

A polymer concentration of 12 wt.% was therefore selected due to its wet spinning compliance across all PVA derivatives. Figure 6 depicts the optical microscopy images of dry wet spun fibres made of native and functionalised PVA. Sample PVA-TB generated homogeneous fibres (F-TB, Figure 6B) with comparable diameter (Ø= 208±18 µm, Figure 6F) with respect to the native PVA variants (Figure 6A). Significantly larger diameter (Ø= 383±58 µm) were measured with UV-cured fibres FA-TB-MA* wet spun from sample PVA-TB-MA and cured in acetone (Figure 6C) compared to both PVA12 and F-TB (Figure 6F).

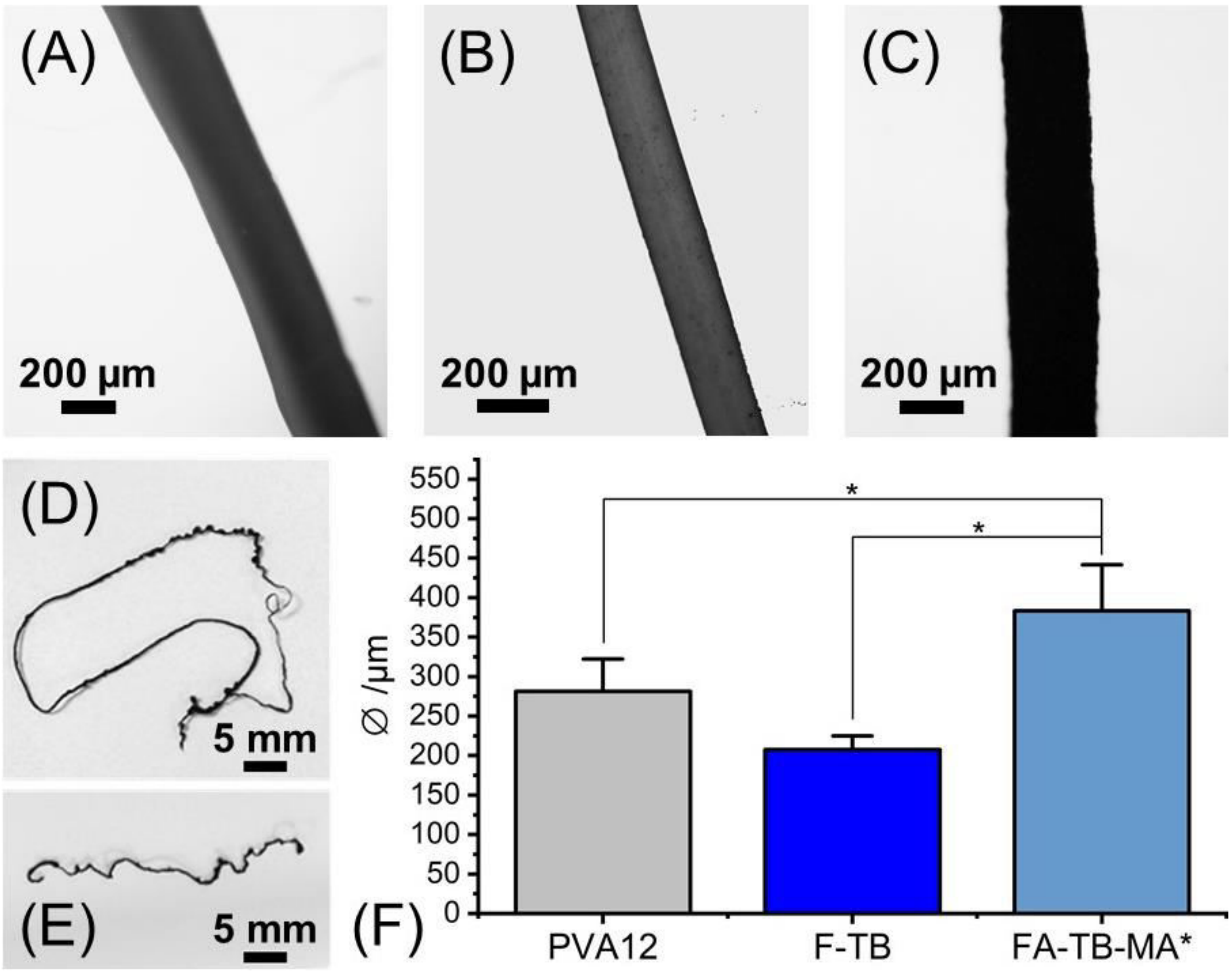

**Figure 6.** Optical microscopy (A-C), photographs (D-E) and fibre diameters (F) of dry wet spun fibres. (A): PVA12, i.e. made of native PVA; (B): F-TB, i.e. made of TB-conjugated product (PVA-TB); (C): FA-TB-MA*, i.e. made of methacrylated, TB-conjugated product (PVA-TB-MA) UV-cured in acetone. All fibres were wet spun from 12 wt.% polymer solutions. (D-E): Photographs of FA-TB-MA* UV-cured in acetone (D) and variant FH-TB-MA* UV-cured in distilled water (E). (F): Fibre diameters (Ø) extracted from optical microscopy images. Data are presented as mean ± standard deviation (n=5, *$p$ <0.05).

When the wet spun polymer PVA-TB-MA was cured in distilled water rather than acetone, shorter fibres were obtained (Figure 6D-E), consequent to the fragility and breakage of the wet material.

The observed increase in fibre diameters revealed by samples FA-TB-MA* compared to PVA12 are attributed to differences in polymer crystallinity, as supported by FTIR spectroscopy (Figure 7). No crystallisation-sensitive band was detected at 1142 $cm^{-1}$ [54,55] in the former compared to the latter samples (Figure 7A and C, and Table S3, Supp. Inf.), consistently with the absence of the same signal in the FTIR spectrum of the corresponding fibre-forming polymer PVA-TB-MA (Figure 2 and Table S1, Supp. Inf.). This observation is also reflected by the lower signal intensity measured at 3270 $cm^{-1}$ in the FTIR spectra of samples PVA-TB-MA and FA-TB-MA* compared to the case of PVA (Table S1, Supp. Inf.) and PVA 12 (Table S3, Supp. Inf.), respectively. Consequently, the introduction of methacrylate residues and UV-induced crosslinks into the PVA backbone is shown to hinder the molecular alignment between fibre-forming polymer chains. The fact that PVA-TB-MA had to be dissolved in a fluorinated solvent such as HFIP, rather than deionised water (as in the case of PVA and PVA-TB), agree with the above observations, aiming to enable the breakdown of hydrogen bonds critical for fibre spinning.

The FTIR spectra of the methacrylated wet spun fibres were further analysed to ascertain the synthesis of a covalent network at the molecular scale following fibre exposure to UV light (Figure 7), as demonstrated with respected polymer solutions (Figure 5B). A lower signal intensity was measured at 1700 $cm^{-1}$ in the spectrum of FA-TB-MA* with respect to the case of FA-TB-MA (Table S3, Supp. Inf.), supporting the consumption of double bonds and the formation of UV-induced covalent crosslinks. Other than that, comparable intensities of the signal at 1630 $cm^{-1}$ (Table S3, Supp. Inf.) were measured in the FTIR spectra of F-TB-MA (Figure 7B) and F-TB-MA* (Figure 7C). Given that this signal was attributed to the C=C groups of TB (Figure 7D), the above observation agrees with the fact that these groups are not affected by the UV-curing process.

Other than the methacrylate PVA fibres, the comparable diameters measured in samples of F-TB, i.e. made of the TB-conjugated polymer, on the one hand, and PVA12, i.e. made of the native polymer, on the other hand, reflected the relatively low content of TB (0.69±0.03 mg per gram of polymer) conjugated to PVA (Figure 4). Consequently, minimal impact on the solution viscosity, and, consequently, on fibre diameter [21,37-39,70], was expected. Together with the aforementioned trends in fibre diameter, the lower fibre length observed with samples of FH-TB-MA* (cured in distilled water) compared to FA-TB-MA* (cured in acetone) was mostly attributed to the specific solvent used for UV-curing.

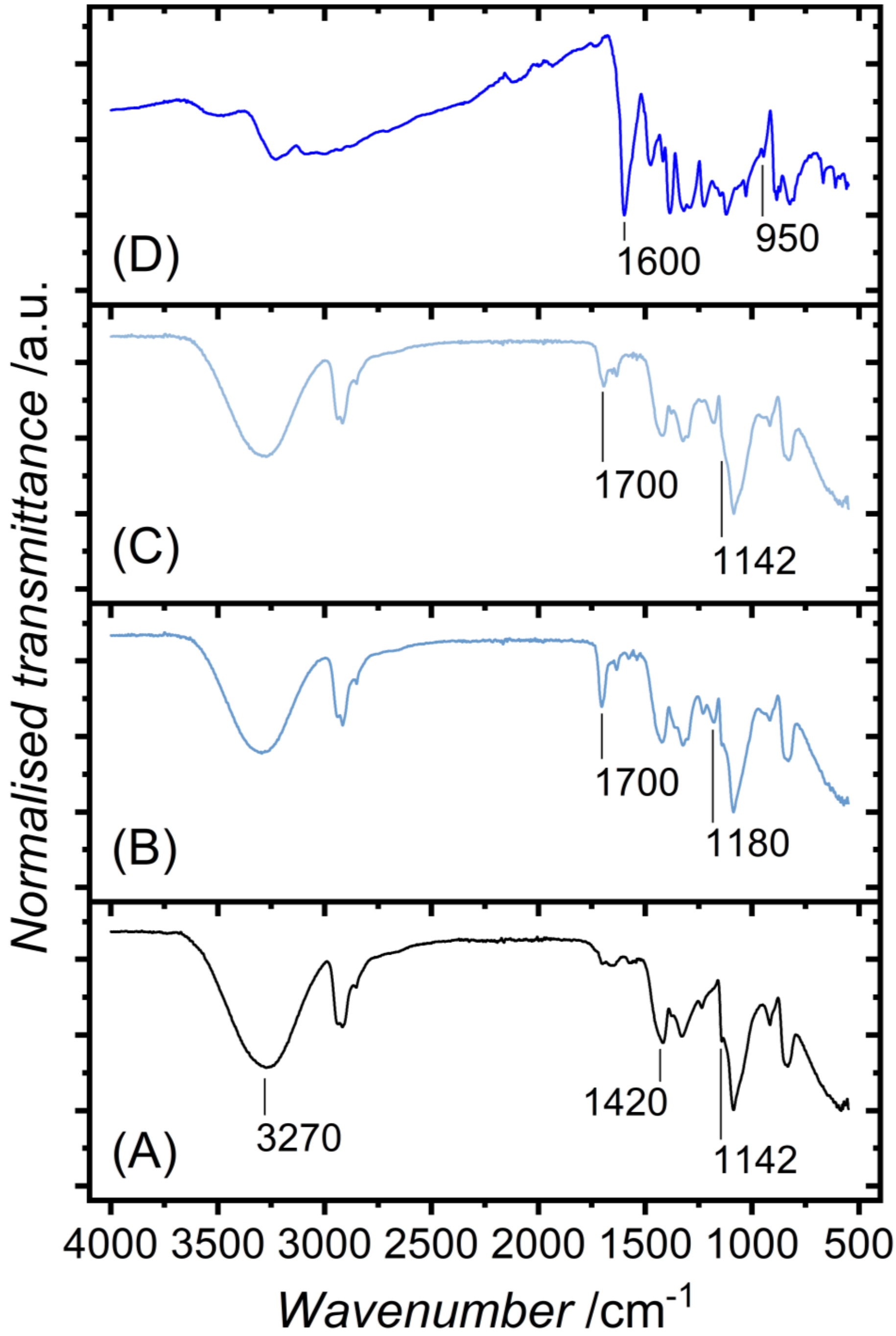

**Figure 7.** FTIR spectra of wet spun fibres (A-C) and TB (D). (A): PVA12 (made of native PVA). (B): F-TB-MA (made of TB-conjugated PVA methacrylate, i.e. PVA-TB-MA). (C): FA-TB-MA* (UV-cured fibre made of PVA-TB-MA).

Although a comparable gel content was measured when the corresponding fibre-forming polymer was cured in the aforementioned solvents (Figure 5A), the incubation in water is associated with risks of fibre swelling, reduced material integrity prior to curing, and brittleness following covalent crosslinking. In contrast, the enhanced integrity observed in fibres cured in acetone agrees with the restricted solubility of the PVA-based fibre in this organic solvent.

### 3.4 Tensile and swelling properties of wet spun fibres

Following characterisation of the microscopic and molecular features of the wet spun fibres, the tensile properties of the dry samples were also explored (Figure S6, Supp. Inf.). More than 20-fold average increase in tensile modulus (*E*: 87±46 → 1835±472 MPa) and more than 4-fold average increase in maximal tensile stress ($\sigma_{max}$: 32±6 →150±41 MPa) were recorded with fibres F-TB with respect to PVA12. This variation in tensile properties suggests the development of secondary interactions between the phenothiazine residues [14-16,67] conjugated to the fibre-forming polymer chains, in line with the drastic reduction in extensibility in the former samples ($\varepsilon_b$: 885±187±185±18%). Similar variations of tensile properties were also measured in fibres PVA10-15, i.e. wet spun from aqueous solutions with increased concentration of PVA, confirming the direct impact of polymer concentration on the molecular entanglements and secondary interactions between fibre-forming polymer chains, as well as on the macroscopic behaviour of the fibres [73,74,75].

UV-curing of the fibres in either acetone (FA-TB-MA*) or distilled water (FH-TB-MA*) induced more than 2.5-fold increase in *E* compared to the corresponding variant made of the native polymer (PVA12). Previous FTIR spectra (Figure 7B-C and Table S3, Supp. Inf.) indicated the formation of a covalently crosslinked molecular network within the UV-cured wet spun fibres. While crosslinking typically enhances mechanical properties, the reduced tensile strength of UV-cured fibres compared to F-TB likely reflects the suppressed polymer crystallinity in the crosslinked state, as indicated by the lack of the crystallisation-sensitive band at 1142 $cm^{-1}$ in the FTIR spectrum of sample FA-TB-MA* (Table S3, Supp. Inf.).

The swelling behaviour of the fibres was subsequently investigated. Samples of PVA12 and FA-TB-MA* revealed a high *SR* (Figure 8A) in both PBS (*SR*= 559±78–699±50 wt.%) and deionised water (*SR*= 717±66–795±19 wt.%), in line with previous reports [76,77,78]. Samples

F-TB could not be tested given their solubility in aqueous environments, reflecting the introduction of crystallisation-hindering phenothiazine residues in respective fibre-forming polymer (i.e. PVA-TB, Table S1, Supp. Inf.) and the lack of UV-induced covalent crosslinks. Likewise, samples FH-TB-MA* could not pursued given the material fragility in the wet state.

Insignificant differences in *SR* and *SI* were measured between samples of PVA12 and FA-TB-MA* in each swelling medium (Figure 8). In contrast to FA-TB-MA*, solvent-equilibrated samples of PVA12 exhibited significantly lower values of *SR* in PBS (*SR*= 559±78 wt.%) compared to deionised water (*SR*= 759±54 wt.%).

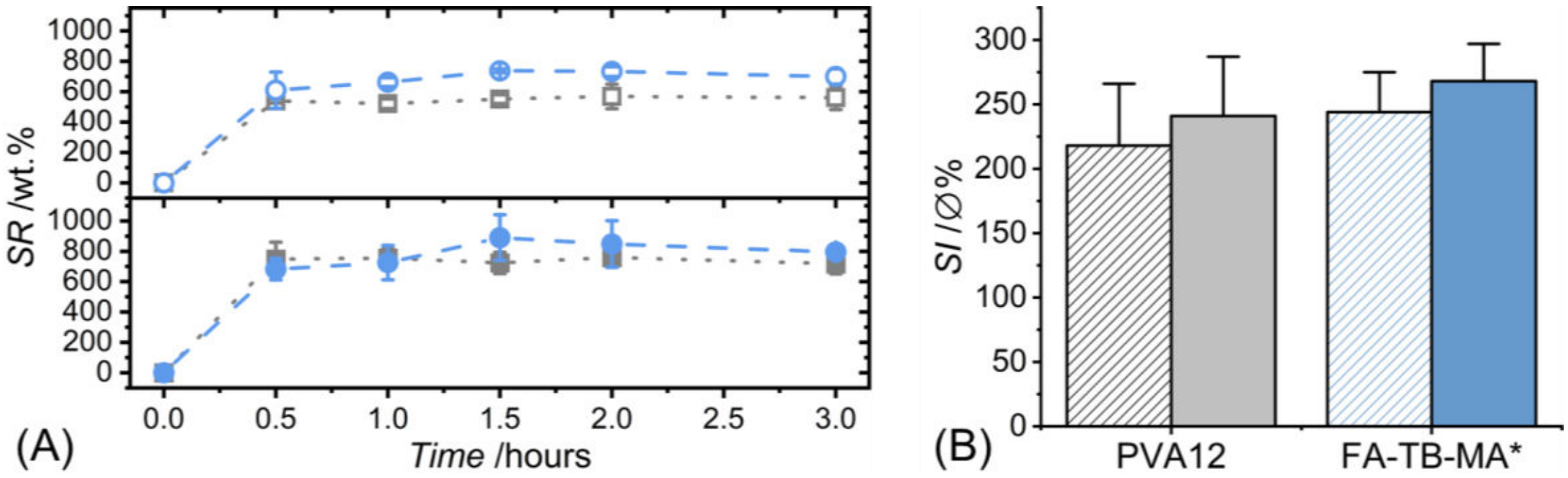

**Figure 8.** Swelling ratio (*SR*, A) and swelling index (*SI*, B) of wet spun fibres. (A): *SR* profile following fibre incubation in either PBS (10 mM, pH 7.4, 37 °C; top) or deionised water (37 °C; bottom). (■,□): PVA12; (●,○): FA-TB-MA*. Data are presented as mean ± standard deviation (n=5). Lines are guidelines to the eye. (B): *SI* measured following fibre equilibration with either PBS (light grey or light blue sparse pattern columns) or deionised water (grey or blue columns). Data are presented as mean + standard deviation (n=5).

At least a 5-fold weight increase was recorded with PVA12 within the first half hour of incubation in both media, with marginal variations observed after this time. These observations point to the effect of the molecular architecture of the fibre-forming polymer on the swelling behaviour of respective wet spun fibres. The consumption of the hydroxyl residues of PVA consequent to the polymer conjugation with TB and MA, together with the introduction of UV-induced covalent crosslinks, is expected to make fibres FA-TB-MA* less sensitive to the presence of salts, as indicated by the comparable values of *SR* and *SI* measured in these samples in both PBS and deionised water. This contrasts the case of PVA12, whereby electrostatic interactions are likely to occur between the ions present in PBS and the hydroxyl residues of native PVA [79], as evidenced by the reduced values of *SR* recorded in PBS compared to deionised water. The presence of the UV-cured polymer network in FA-TB-MA* is also found to delay the swelling rate compared to native PVA fibres, whereby the covalent crosslinks act against the water-induced elongation of polymer chains.

### 3.5 Antibacterial photodynamic capability *in vitro*

Having confirmed the processability of the photodynamic polymer as UV-cured films and textile fibres, the attention moved to exploring respective antibacterial photodynamic capability *in vitro*. The DCFDA fluorometric assay was initially employed to measure the ROS generation following light exposure, and coupled with antibacterial photodynamic tests with *S. aureus* and *P. aeruginosa*, as typical chronic wound bacteria (Figure 9).

The effect of time duration (10, 20, 30 and 60 min) of work light exposure on the generation of ROS was initially investigated with PBS solution controls (Figure S7, Supp. Inf.) supplemented with varying concentrations of TB (0-0.05 mg·mL$^{-1}$). A 60-min light irradiation of low TB concentration controls ([TB] ≤0.01 mg·mL$^{-1}$) revealed the lowest ROS fluorescence signal (2.2±0.1 a.u.), while the highest ROS fluorescence intensity (3.7±0.3 a.u.) was detected after 30-min light exposure. With higher concentrations ([TB] ≥0.02 mg·mL$^{-1}$), an increase in light exposure time, i.e. from 30 to 60 min, proved to be directly related to the ROS fluorescence intensity.

The release of ROS generated from samples of A-TB-MA* was subsequently tested in PBS solution (1 mL) with 30 and 60 min of light exposure (Figure 9A). The weight of the film proved to have a direct impact on the intensity of the fluorescence signal with both light exposure durations, in line with the variation in the amount of TB residues conjugated to the polymer (Figure 4B). In contrast to the sample weight, the duration of light exposure did not play a significant role on the extent of ROS generation, so that similar signal intensities were measured after 30-min and 60-min of light exposure, across all the sample weights tested.

Other than the work light, it was also of interest to investigate whether the UV-curing treatment used to crosslink films made of PVA-TB-MA yielded any detectable ROS fluorescence intensity.

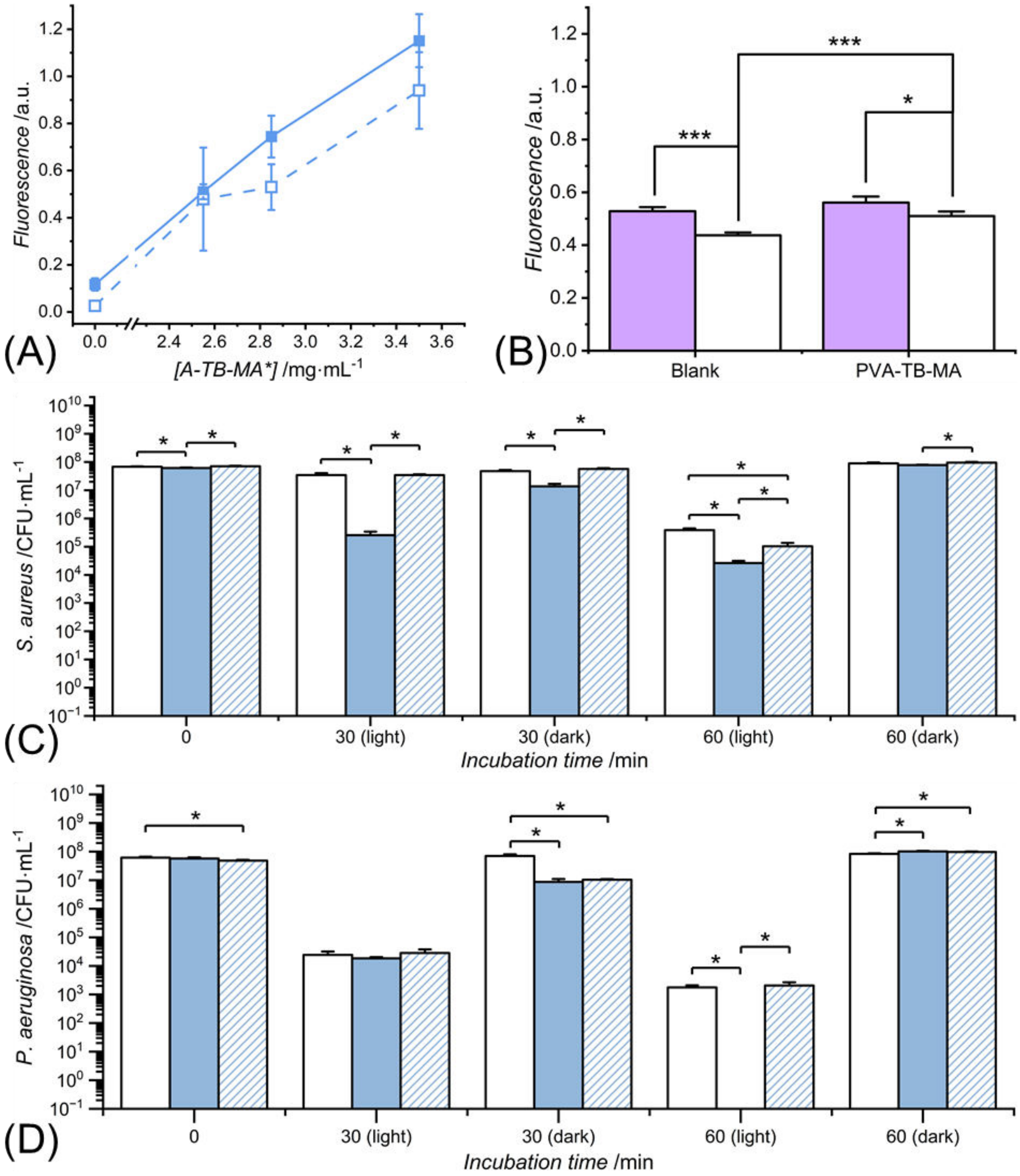

**Figure 9.** Photodynamic tests with work light and UV light. (A): ROS assay fluorescence intensity measured on UV-cured films of A-TB-MA* following incubation in 2 mL PBS and 30 min (—■—) and 60-min (--□--) work light exposure. Lines are guideline for the eye. (B): ROS assay fluorescence intensity measured following 30-min UV light exposure (■) and 60-min work light exposure (□) of film-forming photoactive sample PVA-TB-MA (5.1 mg in 2 ml of acetone) and sample-free acetone control. (C-D): Antibacterial effect against *S. aureus* (C) and *P. aeruginosa* (D) exhibited by UV-cured films (light blue columns) and UV-cured fibres (light blue patterns) following either 30-60 min–light exposure or 30-60 min–dark incubation. White columns refer to the sample-free bacteria controls. Data are presented as mean ± standard deviation (n=3); * $p$ <0.05, *** $p$ <0.001.

After 30-min of UV light exposure in AcOH, the signal revealed by the corresponding films was comparable to the one detected with the sample-free control of AcOH after the same duration of UV exposure (Figure 9B). On the other hand, a significantly increased ROS signal was detected following 60-min work light irradiation of PVA-TB-MA in AcOH compared to the sample-free control. The above observations therefore indicate that it is possible to expose the PVA derivative to UV light with minimal risks of ROS generation; they also demonstrate the photodynamic response of the PVA derivative to work light exposure. In this context, the intensity of the ROS fluorescence signal (0.51±0.02 a.u.) recorded with sample of PVA-TB-MA following 60-min work light irradiation in acetone (Figure 9B) proved to be comparable to the intensity (0.48±0.22 a.u.) measured with the UV-cured film A-TB-MA* irradiated with work light for the same time duration in PBS (Figure 9A). This observation further supports the minimal impact of the UV-curing step and incubation solvent on the photodynamic capability of the film-forming PVA derivative.

Following demonstration of ROS generation under work light, the antibacterial effects of A-TB-MA* against planktonic *S. aureus* (Figure 9C) and *P. aeruginosa* (Figure 9D) were evaluated by colony forming unit (CFU) assay with varied work light exposure durations. Although the killing of *S. aureus* and *P. aeruginosa* by photosensitisers belonging to different antibacterial dye compounds and formulations have been reported [22,23,27,29,80], no study has disclosed the inactivation of both strains by UV-cured TB-conjugated films and fibres upon light irradiation. Significant antibacterial effects on *S. aureus* were triggered by both films and fibres following 60-min light exposure, reflecting a mean bacterial viability reduction of 1.17 log (93%) and 0.57 log (73%) reduction, respectively. With 30-min light exposure, films displayed tremendous mean bacterial reduction of 2.13 log (99%) reduction, while an insignificant antibacterial effect was measured with respective fibres, likely due to the relatively smaller size of the latter compared to the former samples. Other than *S. aureus*, no significant differences were found in *P. aeruginosa* viability after 30-min light exposure of either films or fibres (Figure 9D). Remarkably, complete (100%) bacterial eradication was observed following 60-min light exposure of the films, while no significant effect was observed with the fibres. These results further support the influence of the material geometry (fibre vs. film) on the antibacterial photodynamic effect. Interestingly, both bacterial strains presented colony reduction in 30-min dark conditions rather than in 60-min dark conditions. This observation may be attributed to the initial effect of the cationic character of PVA-TB-MA, consequent to the introduction of the positively charged TB residues, as known for carbohydrate polymers with a similar molecular backbone [81,82].

Qualitative SEM images of the bacterial cell morphology on films and fibres (Figure S8, Supp. Inf.) reflected the quantitative bacterial viability CFU values. Compared to intact *P. aeruginosa* or *S. aureus*, bacterial colonies displayed no morphology difference following both 30- and 60-min dark incubation, whereas significant colonies reduction was observed following incubation with comparable irradiation time.

Overall, both UV-cured samples presented a significant antibacterial effect depending on the time of light irradiation and the material geometry (film vs. fibre) consequent to the ROS generation triggered by the work light-irradiated TB residues immobilised on the polymer. The work light-induced photodynamic capability of this material can therefore be a promising alternative to tackling bacterial infections in chronic wounds.

### 3.6 Cellular tolerability *in vitro*

Having demonstrated the antibacterial photodynamic capability of the UV-cured film and fibre, the attention moved to investigating the tolerability of these materials with L929 mouse fibroblasts following light exposure. Extract and contact tests were therefore performed, and the cellular metabolic activity (Figure 10) and viability (Figure 11) monitored over 72 hours following an initial 60-min exposure to work light.

The relative metabolic activity of cells exposed to film extracts was found to slightly increase from 82 ± 5 % to 93 ± 5 %, as the culture time was increased from 24 to 72 hours, respectively (Figure 10A). These values were comparable to the ones measured following cell exposure to fibre extracts, whereby no significant difference was observed during the selected cell culture time points. Qualitative observations of cell viability from live/dead staining images agreed with previous quantitative metabolic activity measurements. A predominance of live cells (green), compared to dead cells (red), was observed across all extracts tested (Figure 11A-F). The number of live cells was found to increase with the cell culture time, indicating increased cellular metabolic activity and proliferation, minimal leaching of cytotoxic compounds, and, ultimately, cellular tolerability of both film and fibre extracts.

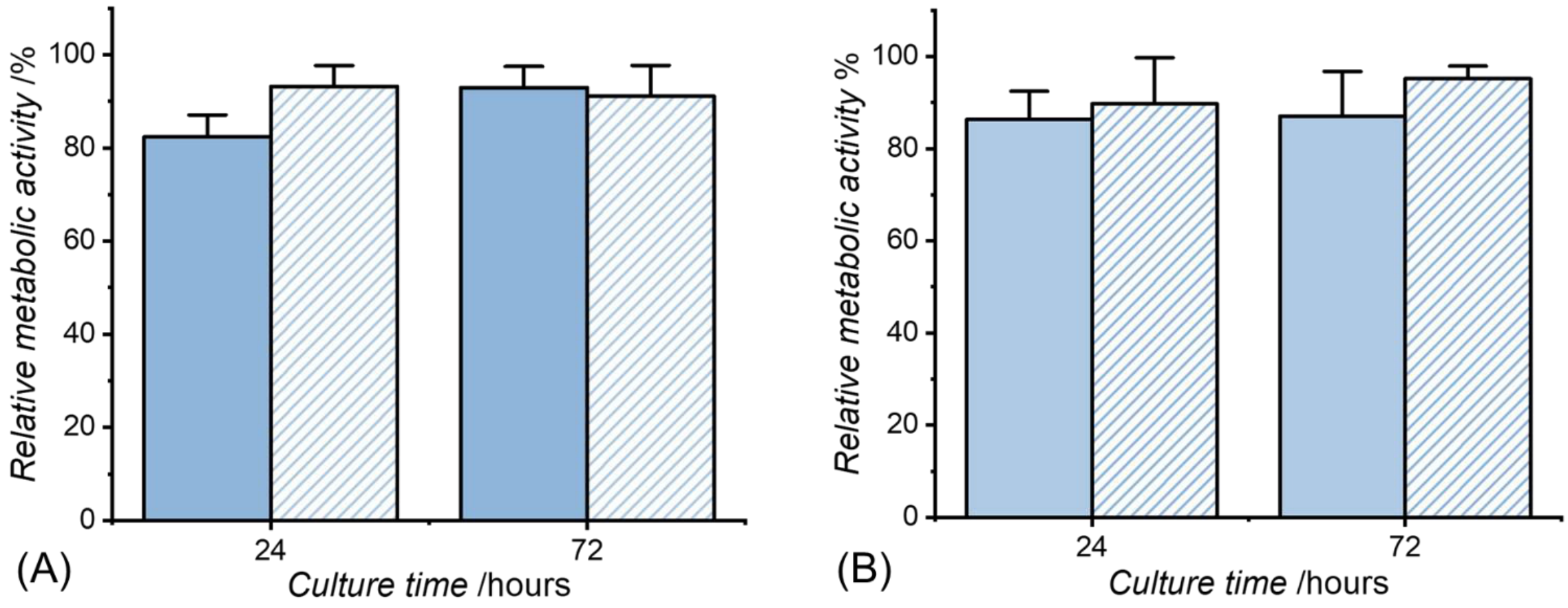

**Figure 10.** Relative metabolic activity of L929 fibroblasts measured via Alamar Blue assay following 24- and 72-hour cell culture. Extracts (A) of UV-cured films (A-TB-MA*) and fibres (FA-TB-MA*) were tested together with contact tests (B) with the corresponding samples. Blue and sparse columns refer to samples and extracts of A-TB-MA* and FA-TB-MA*, respectively. Data are presented as mean ± standard deviation (n=3) relative to the cellular metabolic activity measured on tissue culture plastic control at each culture time point.

In addition to monitoring the cellular tolerability of UV-cured sample extracts, a direct contact test was conducted, whereby L929 mouse fibroblasts were seeded into wells containing either films A-TB-MA* or fibres FA-TB-MA*, alongside tissue culture plastic control. The relative metabolic activity of cells cultured onto films remained fairly constant across the two culture time points, reaching 86 ± 6 % after 24 hours and 87 ± 10 % after 72 hours (Figure 10B). Similar trends were observed following contact tests with the fibres, whereby the relative metabolic activity of cells was recorded at 90 ± 10 % after 24 hours and 95 ± 3 % after 72 hours. The slightly lower, though insignificantly different, values measured with the films compared to the fibres may be attributed to the different geometry of the samples, as discussed with previous antibacterial photodynamic tests (Figure 9C-D). Indeed, the UV-cured films were made with a diameter of about 8 mm, in contrast to the micrometre scale (Ø= 383±58 μm) of the corresponding fibres (Figure 6F). Qualitative cell viability observations from live/dead staining images of cells cultured in contact with the fibres (Figure 11G-L) corroborated the aforementioned metabolic activity data, showing a predominance of live cells (green) compared to dead cells (red) across all extracts tested. As observed with the fibre extracts, the number of vital cells was found to increase from 24 to 72 hours, indicating cell proliferation and providing further evidence of cellular tolerability.

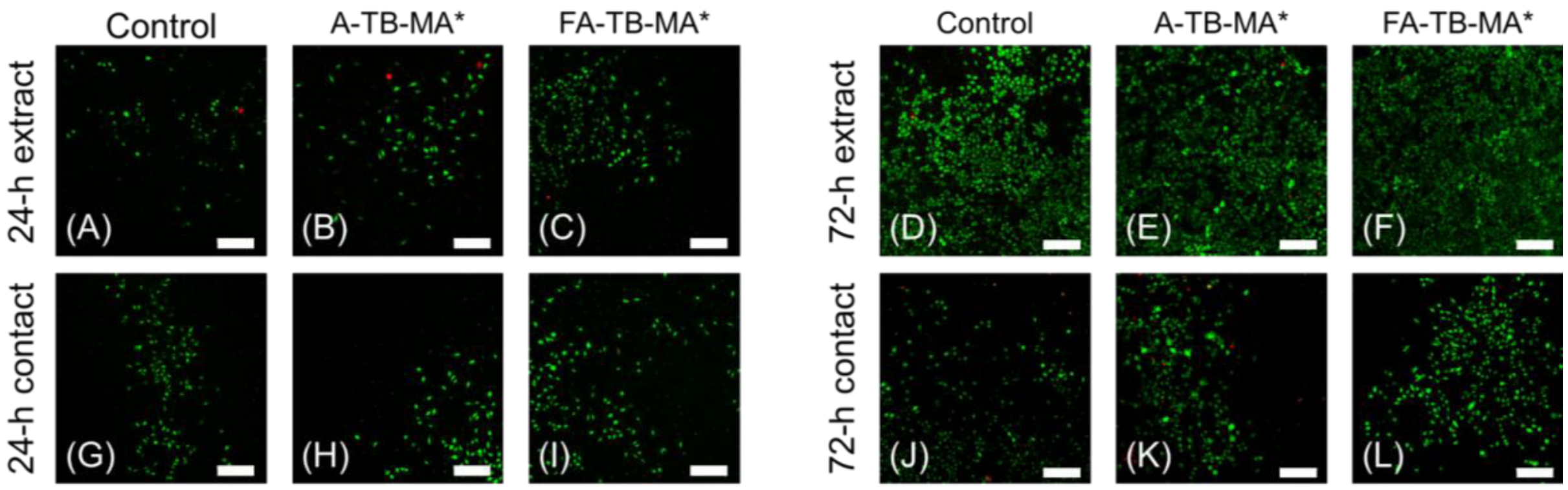

**Figure 11.** Live/dead microscopy images of L929 fibroblasts captured following 24-hour (A-C, G-I) and 72-hour (D-F, J-L) culture with DMEM control (A, D) and extracts of film A-TB-MA* (B, E) and fibre FA-TB-MA* (C, F). Tissue culture plastic control (G, J), and the corresponding film (H, K) and fibre (I-L) were also tested in contact tests. Scale bars: 200 μm.

Taken together, these results highlight that the UV-cured samples of film and fibre did not trigger any cytotoxic effects to L929 fibroblasts following both 60-min work light exposure and 72-hour cell culture, while still enabling significant antimicrobial photodynamic effects (Figure 9). This targeted photodynamic capability is key aiming to accomplish antibiotic-free infection control while ensuring minimal toxicity to endogenous mammalian cells.

### 3.7 Integration in a photodynamic wound dressing

Given the cellular tolerability and photodynamic capability of the films and fibres developed in this study, it was of interest to investigate how the PVA derivative could be integrated into therapeutic wound dressings, aiming to generate both wound healing and antibacterial capabilities. The advanced wound dressings employed in the clinic are typically made of multiple layers, whereby each layer is equipped with a specific and independent functionality. Following this design approach, it was hypothesised that the PVA derivative could be delivered as a wound dressing antibacterial coating. Given their significantly higher antibacterial effect, the UV-cured film was preferred to the fibre format to demonstrate integration with the wound dressing. The presence of the methacrylate residues on the polymer backbone of the PVA derivative was leveraged aiming to generate photoinduced covalent linkages at the interface with the underlying wound dressing (Figure S9, Supp. Inf.), aiming to minimise risks of coating delamination following contact with the wound environment and liquid uptake. A collagen-based wound dressing made of a photoactive collagen precursor and with demonstrated wound healing capability *in vivo* [46,47] was therefore employed to demonstrate this approach. The dressing precursor, namely 4VBC, bears photoactive residues along its collagen backbone, enabling the synthesis of covalent linkages between the collagen layer and the PVA coating after UV curing.

Figure 12A depicts the wound dressing composite, coded as C-PVA*, that was successfully obtained via sequential casting of the I2929-supplemented solutions of 4VBC and PVA-TB-MA, prior to single UV exposure.

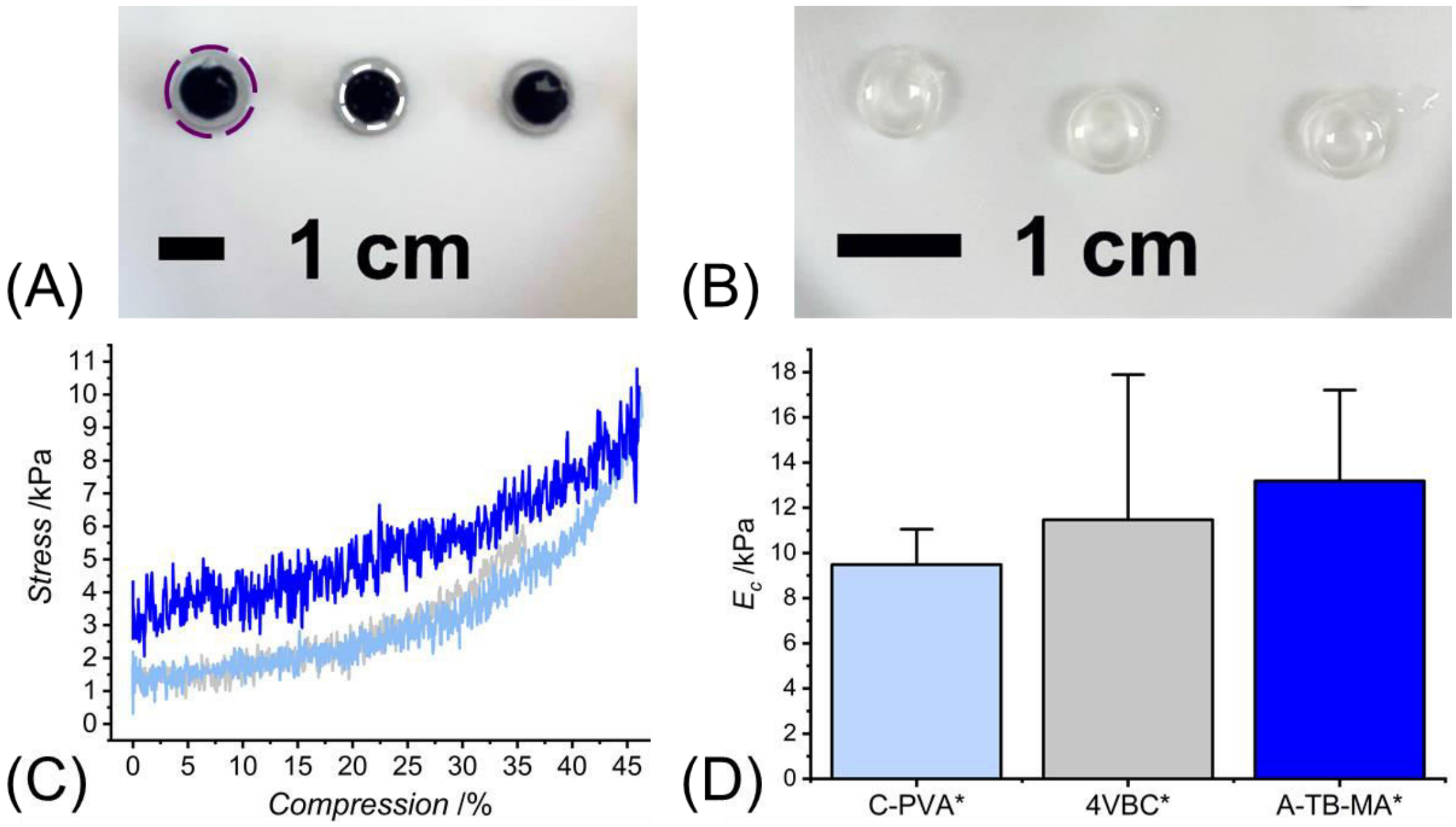

**Figure 12.** Integration of the photodynamic PVA derivative in a UV-cured bilayer composite (C-PVA*). (A): photographs of C-PVA* (A) made of the collagen-based bottom layer (purple dashed line) and photodynamic top layer (white dashed line). (B): photographs of the UV-cured PVA-free control of 4VBC* (B). (C-D): stress-compression curves (C) and compression moduli (D) of composite C-PVA* (light blue), PVA-free control of 4VBC* (light grey) and 4VBC-free control of A-TB-MA* (blue). Data are presented as mean ± standard deviation (n=8).

The presence of the photodynamic PVA coating in C-PVA* was clearly visible, due to its distinct TB-associated blue colour, in contrast to the transparent appearance (Figure 12B) of the collagen-based layer (coded as 4VBC*). Compression testing of the freshly synthesised sample of C-PVA* was successfully carried out with no detected structural delamination. Comparable stress-compression curves (Figure 12C) and compression moduli (Figure 12D) were measured with respect to the case of each composite constituent tested in the same conditions. This observation provides indirect evidence of the nanoscale integration between the two layers and indicates insignificantly different compression properties between the samples of 4VBC* and A-TB-MA*. Consequently, it is possible to coat the collagen-based dressing with the photodynamic film with minimal alteration of the native mechanical properties.

In contrast to the single-cured two-layer composite, complete structural delamination was observed during physical handling of the composite controls, which were prepared by sequential casting and UV-curing of the solutions of collagen precursor and PVA-TB-MA. These observations suggest the lack of covalent linkages at the dressing interface with the coating, which was otherwise observed through single UV exposure of the two photoactive polymer solutions. The aforementioned control experiments also suggest a limited mechanical entrapment of the PVA phase in the collagen layer, likely attributed to the relatively high viscosity of the collagen solution prior to curing. Further studies will therefore focus to investigate whether the aforementioned nanoscale integration of the photodynamic PVA coating can be achieved with other wound dressings of different chemical compositions, and whether the presence of the composite structure has any impact of the photodynamic response of the PVA coating.

## Conclusions

This study represents a significant advancement in the field of antibiotic-free antimicrobial materials by introducing a multifunctional polyvinyl alcohol (PVA) derivative with staining-free photodynamic capability for use in advanced antibiotic-free wound dressings. Conjugation of PVA with both toluidine blue (0.69±0.03–0.81±0.05 mg per gram of polymer) and methacrylate residues (up to 7 mol.% of vinyl alcohol repeat units) is successfully demonstrated via UV-Vis spectroscopy and proton nuclear magnetic resonance spectroscopy, respectively. Consequently, suppressed toluidine blue release (0.37±0.05 µg) and toluidine blue-induced staining are observed over a 96-hour incubation in phosphate buffered saline, together with the viability of L929 murine fibroblasts following 60-min photodynamic material exposure and 72-hour cell culture. The molecular architecture of the PVA derivative ensures UV-induced material integrity in near-physiological conditions, and polymer processability into multiple wound dressing-compliant material formats, i.e. cast film (*shear modulus*: 8-12 kPa; *loss modulus*: 1-5 kPa), wet spun fibre (*diameter*: 208±18–383±58 µm), and nanoscale integrated coating (*compression modulus*: 13±4 kPa). Rheology, Fourier Transform Infrared (FTIR) spectroscopy and gel content measurements support the formation of a covalently crosslinked molecular network in the UV-cured state, yielding a convenient strategy for the development of durable photodynamic composite dressings (*compression modulus:* 9±2 kPa). The synthesis of the covalently-crosslinked network in the UV-cured fibres is found to inhibit the polymer crystallinity of native PVA, as indicated by the suppression of respective crystallisation-sensitive band (1142 $cm^{-1}$) in the FTIR spectra of the UV-cured fibres, and to provide an additional dimension for the control of tensile and swelling properties. A fluorometric assay confirms the ability of the PVA derivative and respective UV-cured network to generate reactive oxygen species (ROS) upon work light, and not UV light, exposure. This indicates minimal impact of the UV curing step and light exposure time (30-60 min) on the photodynamic response, and a direct relationship of the latter with the sample weight investigated (2.55-3.5 mg). Accordingly, both UV-cured films and UV-cured fibres demonstrate significant antibacterial effects against Gram-positive *S. aureus* (0.57-2.13 log reduction). Complete eradiation of Gram-negative *P. aeruginosa* is also observed following 60-min film exposure to work light, whereby the significantly higher bacterial inactivation observed with the films with respect to the fibres likely reflects differences in geometry between the two samples. Collectively, the above findings indicate that this PVA derivative offers a promising alternative to traditional antibiotics and state-of-the-art antimicrobial wound dressings, addressing the growing issue of antibiotic resistance with a scalable and regulatory-friendly medical device approach. The potential applications of this technology extend beyond wound dressings to encompass various healthcare technologies enabling controlled, localised therapeutic activity, from infection control up to minimally invasive anticancer therapy.

**Conflicts of interest**

Giuseppe Tronci (GT), Man Li and Charles Brooker are named inventors on a patent application related to the design of photodynamic polymers. GT has equity in and serves on the board of directors of HYFACOL Limited.

**Acknowledgements**

The authors gratefully acknowledge financial support from the Clothworkers' Company (London, UK), the Grow MedTech's Proof of Feasibility programme supported by UKRI Research England's Connecting Capability Fund [project code: CCF11-7795], BBSRC (Biotechnology and Biological Sciences Research Council) under grant BB/X011631/1, EPSRC (Engineering and Physical Sciences Research Council) under grant EP/P027687/1 and project reference 2111340, and the Royal Society International Exchanges Cost Share Award, UK (IEC\NSFC\223289). The technical assistance provided by Michael Brookes is gratefully acknowledged.

**Data availability statement**

Data will be made available on request.

**References**

---

## Supporting Information

**Photodynamic, UV-curable and fibre-forming polyvinyl alcohol derivative with broad processability and staining-free antibacterial capability**

Man Li,[1,2] Charles Brooker,[1,2] Rucha Ambike,[1] Ziyu Gao,[2,3] Paul Thornton,[1,3] Thuy Do,[2] Giuseppe Tronci[1,2,*]

[1] Clothworkers' Centre for Textile Materials Innovation for Healthcare, Leeds Institute of Textiles and Colour, School of Design, University of Leeds, Leeds, LS2 9JT, United Kingdom

[2] School of Dentistry, St. James's University Hospital, University of Leeds, Leeds, LS9 7TF, United Kingdom

[3] School of Chemistry, University of Leeds, Leeds, LS9 7TF, United Kingdom

[*] Email correspondence: g.tronci@leeds.ac.uk

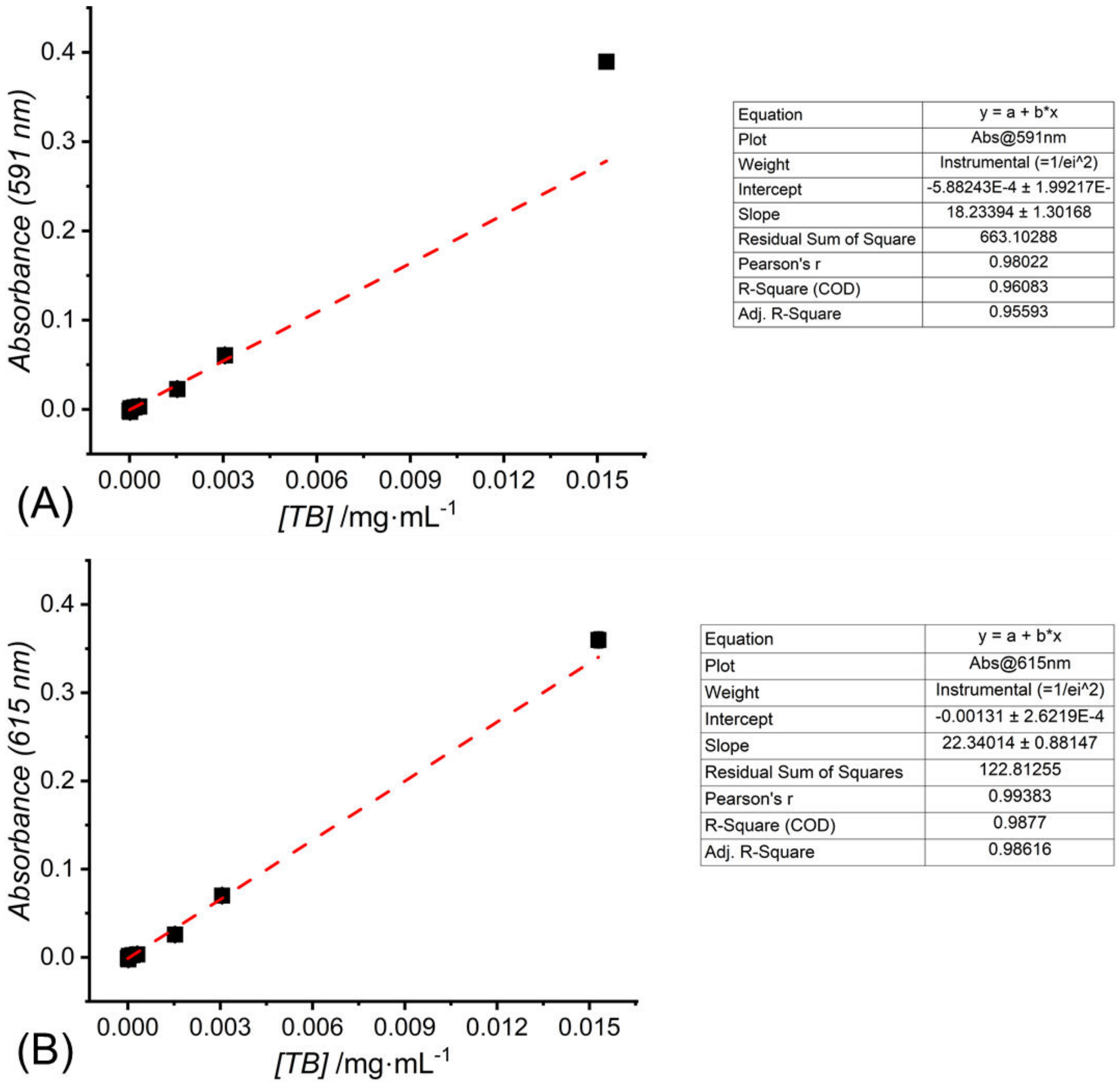

**Figure S1.** UV-Vis calibration curves built with aqueous solutions ($H_2O$, 25 °C) supplemented with varied concentrations of TB. The absorbance values recorded at 591 nm (A) and 615 nm (B) are presented as mean ± standard deviation (n=2). The red line describes the linear fit of the mean values according to the equation provided in each plot.

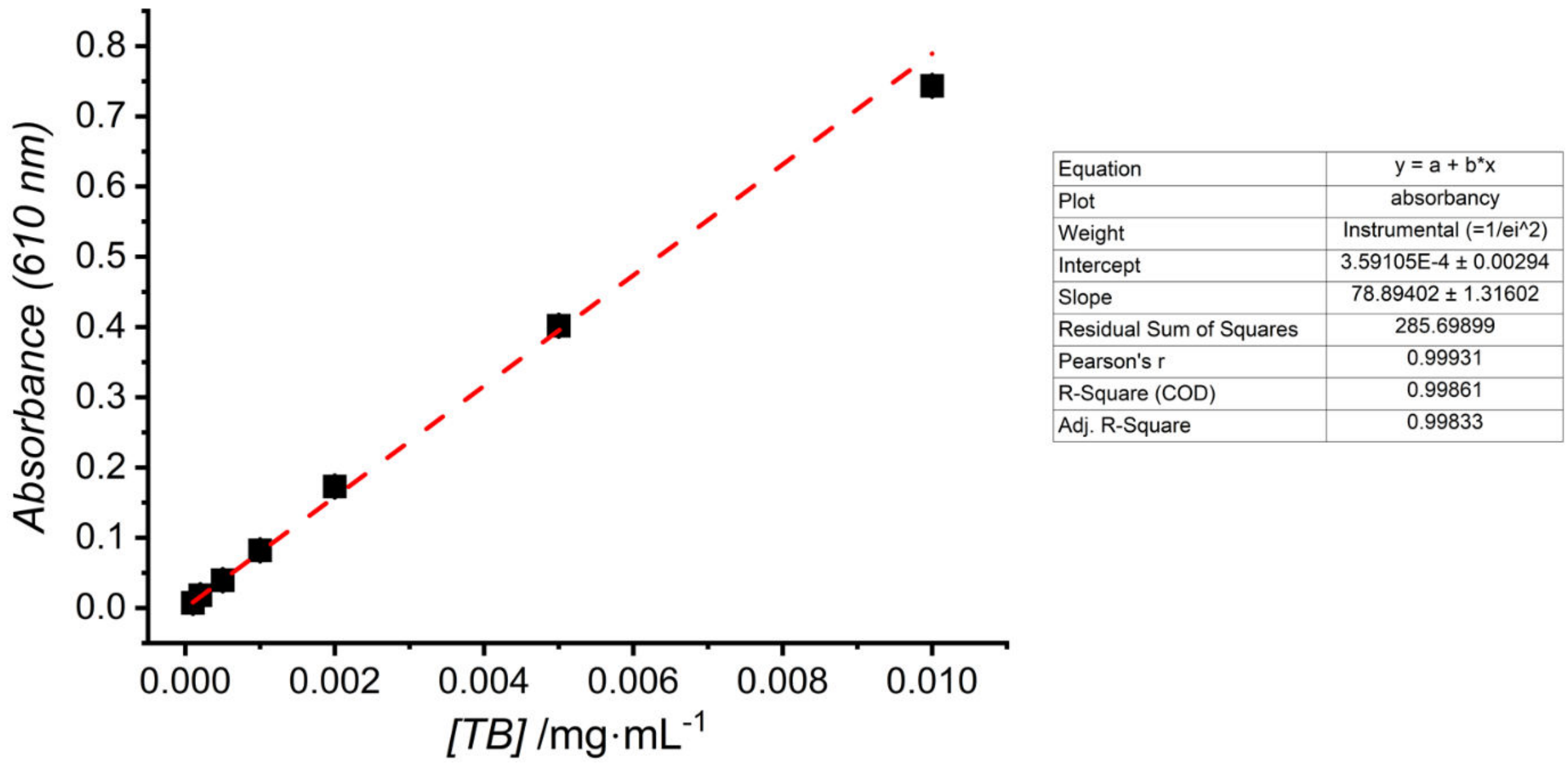

**Figure S2.** UV-Vis calibration curves built with PBS solutions (10 mM, pH 7.4, 25 °C) supplemented with varied amount of TB. The absorbance recorded at 610 nm is presented as mean ± standard deviation (n=2). The red line describes the linear fit of the mean values according to the equation provided in the plot.

**Table S1.** Integrals of FTIR signals revealed by native PVA as well as TB- and MA-reacted products. n.o.: not observed.

| Band /cm$^{-1}$ | PVA | PVA-TB | PVA-TB-MA |
|---|---|---|---|
| 1087 | 28 | 66 | 42 |
| 1142 | 3 | n.o. | n.o. |
| 3270 | 204 | 222 | 167 |

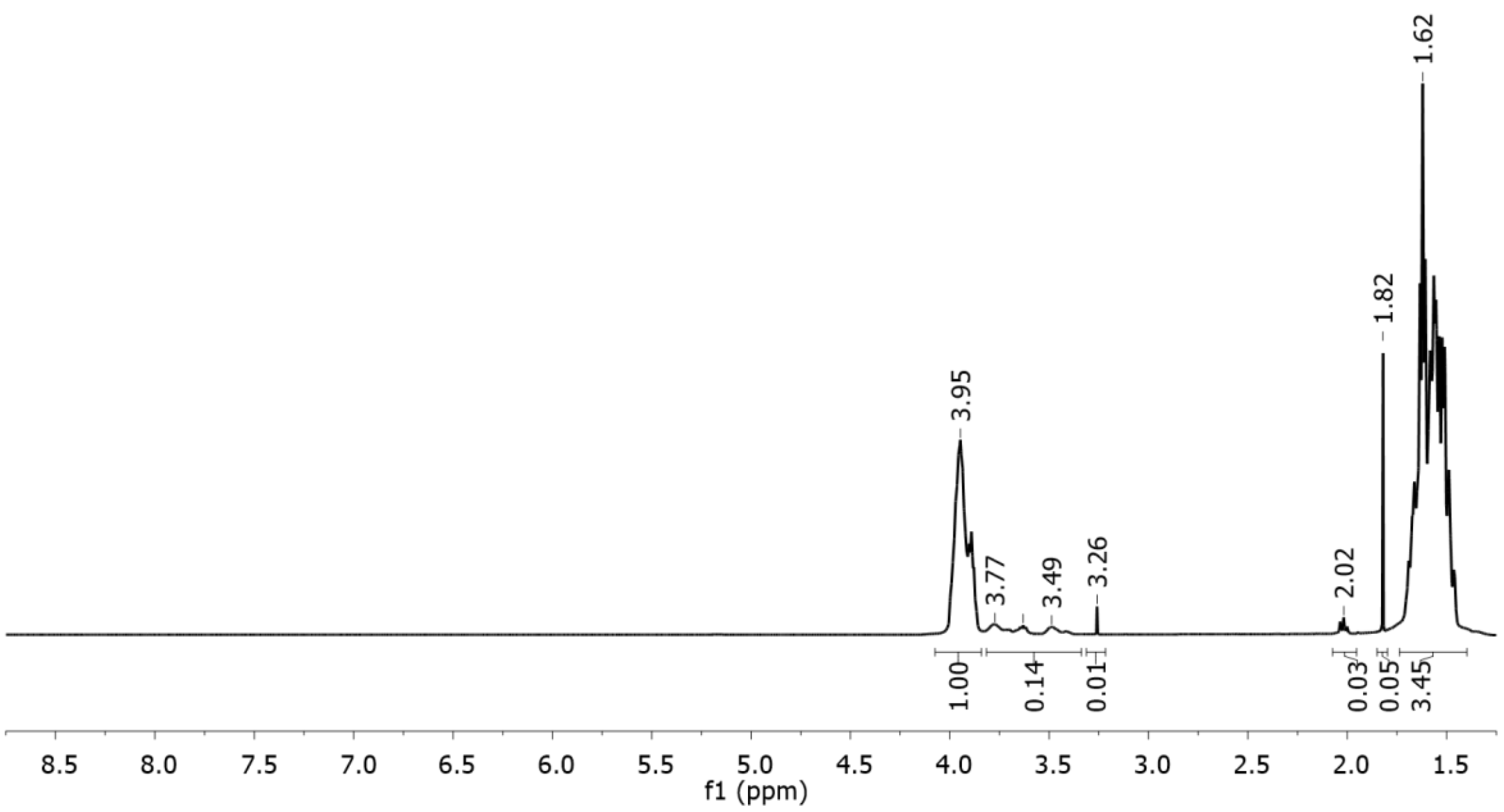

**Figure S3.** $^1$H-NMR spectra (500 MHz, $D_2O$, 25 °C) of native PVA. The signals of the relevant peaks are assigned for the determination of the chemical composition.

**Table S2.** Quantification of PVA grafting via $^1$H-NMR and signal integration. The degree of tosylation was quantified through integral ratios of tosyl-related signals (either at 7.3-7.6 ppm or 2.3 ppm) to PVA signals (either at 1.6 ppm$^a$ or 3.9 ppm$^b$). The degree of methacrylation was calculated from the integral ratios of the methacrylate signals (either at 5.6-6.1 ppm or 1.9 ppm) to PVA signals (either at 1.6 ppm$^a$ or 3.9 ppm$^b$). n.o.: not observed.

| **Sample ID** | **Tosylation /mol.%** | | **Methacrylation /mol.%** | |
|---|---|---|---|---|
| | **7.3-7.6 ppm** | **2.3 ppm** | **5.6-6.1 ppm** | **1.9 ppm** |
| PVA-Ts | 7$^a$; 11$^b$ | 11$^a$; 17$^b$ | n.o. | n.o. |
| PVA-TB-MA | n.o. | n.o. | 4$^a$; 7$^b$ | 4$^a$; 7$^b$ |
| PVA-TB-MA (control) | 0.1$^a$; 0.2$^b$ | 0.2$^a$; 0.3$^b$ | 4$^a$; 7$^b$ | 4$^a$; 7$^b$ |

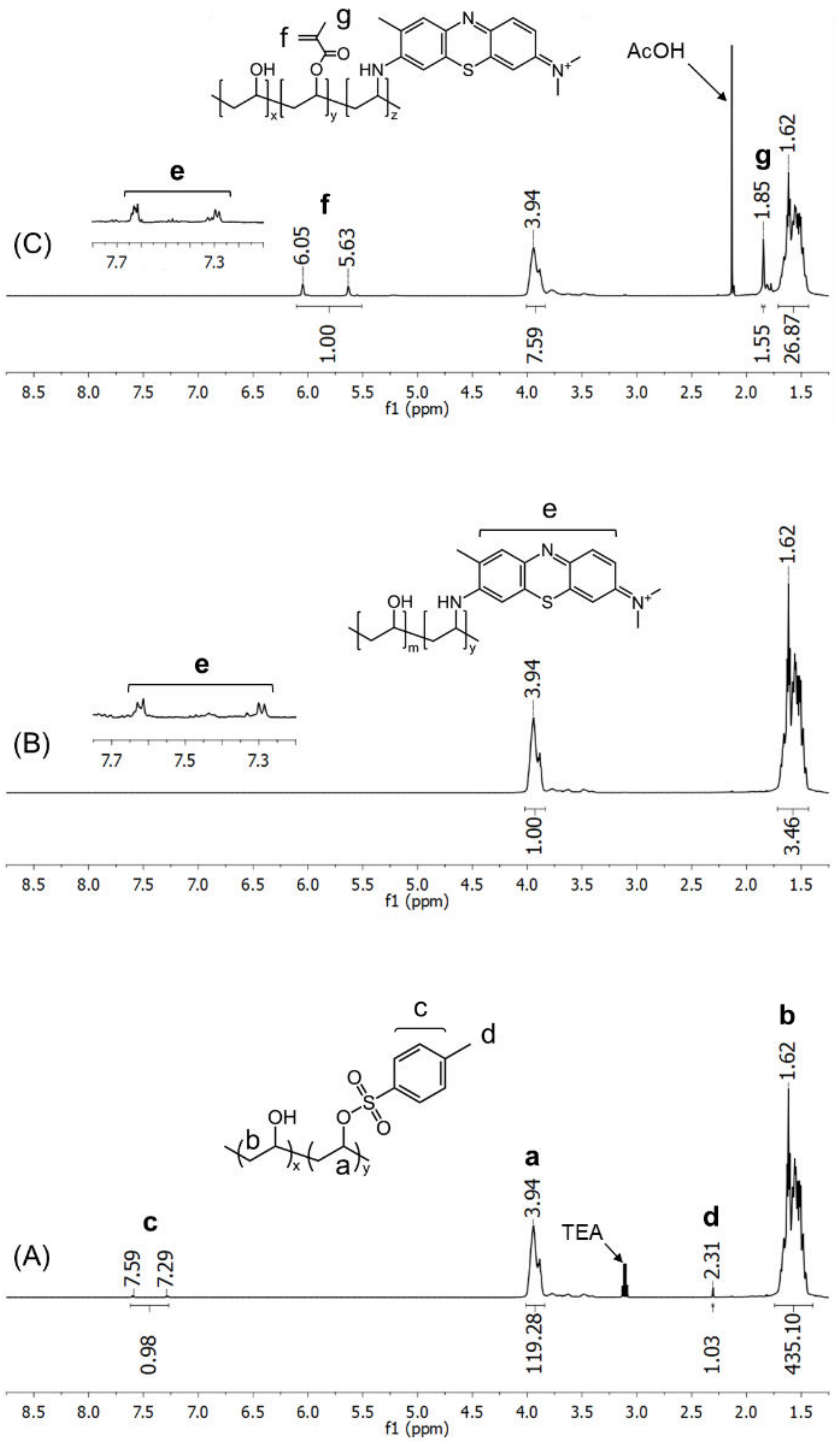

**Figure S4.** $^1$H-NMR spectra ($D_2O$, 25 °C) of PVA-Ts (A), PVA-TB (B) and PVA-TB-MA (C), revealing the detection of tosylate ($[Ts]\cdot[OH]^{-1}$= 0.1, A), phenothiazine aromatic (zoomed-in area, B), and methacrylate signals (C, $[MA]:[OH]^{-1}$= 0.5), respectively. The complete elimination of the tosylate signals is also observed in (B).

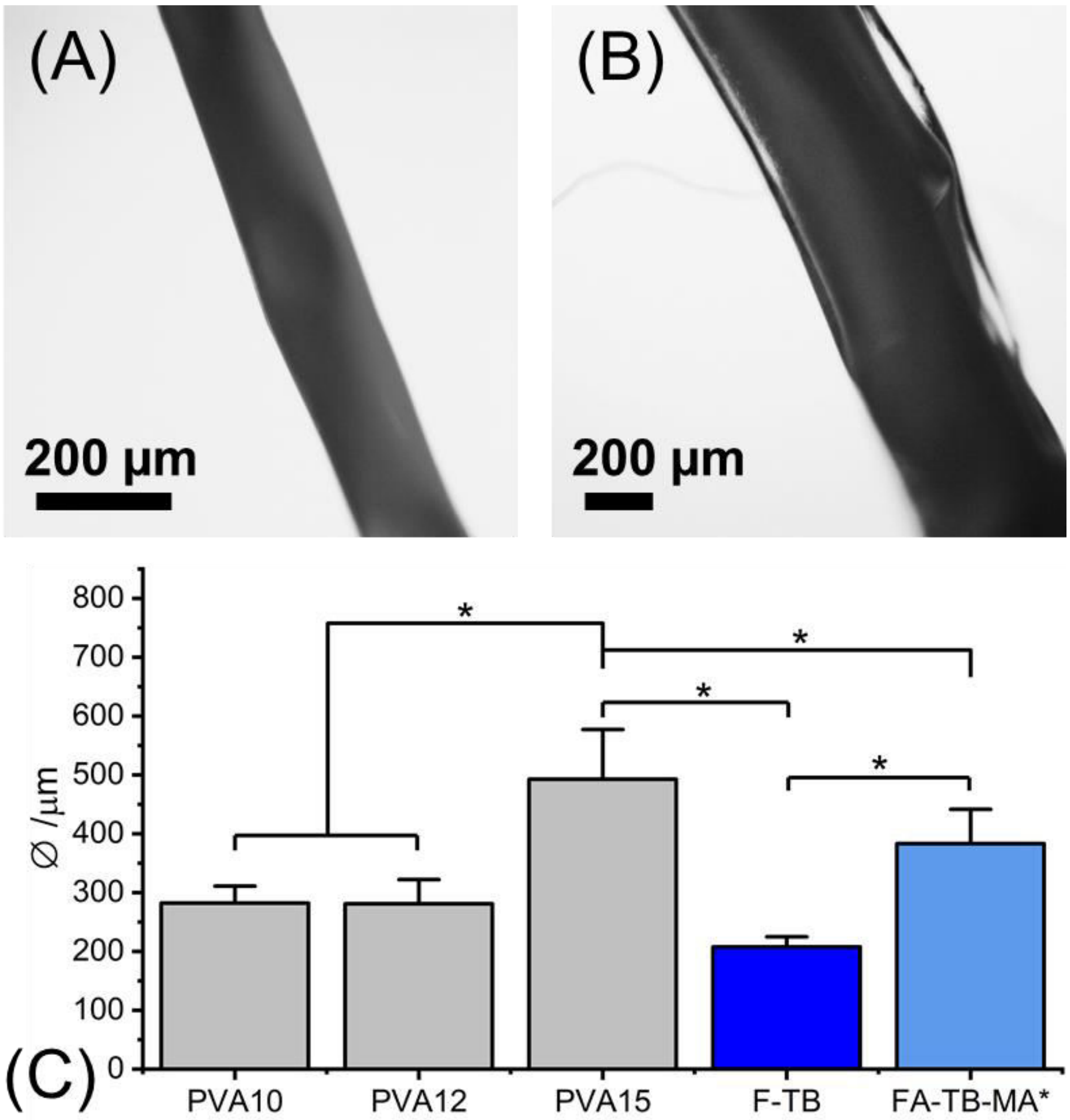

**Figure S5.** Effect of polymer dope concentration on appearance of wet spun fibres made of native PVA. (A-B): Optical microscopy images of fibre PVA10 (A) and PVA15 (B). (C): Comparison of fibre diameter across the whole set of fibre samples. Data are presented as mean ± standard deviation (n=5, *$p$< 0.05).

**Table S3.** FTIR Integrals recorded in fibres made of native (PVA12) and methacrylated (F-TB-MA) PVA, as well as methacrylated PVA fibres following UV curing in acetone (FA-TB-MA*). n.o.: not observed.

| Band /cm$^{-1}$ | PVA12 | F-TB-MA | FA-TB-MA* |
|---|---|---|---|
| 1142 | 1 | 1 | n.o. |
| 1630 | 0 | 1 | 1 |
| 1700 | 1 | 9 | 4 |
| 3270 | 228 | 189 | 193 |

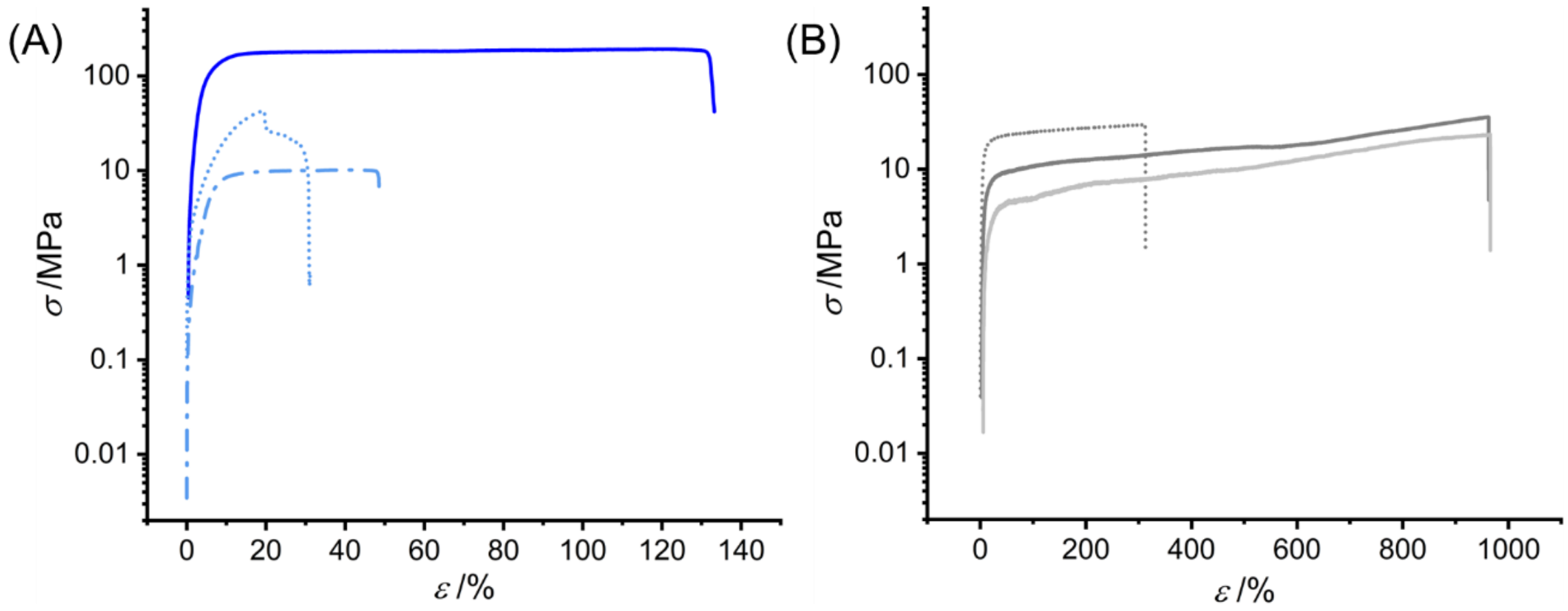

(C)

| Fibre ID | PVA10 | PVA12 | PVA15 | F-TB | FA-TB-MA* | FH-TB-MA* |
|---|---|---|---|---|---|---|
| $E$ /MPa | 32 ± 16 * | 87 ± 46 * | 159 ± 44 * | 1835 ± 472 # | 227 ± 90 * | 222 ± 74 * |
| $\sigma_{max}$ /MPa | 20 ± 3 * | 32 ± 6 * | 44 ± 14 * | 150 ± 41 # | 11 ± 1 * | 31 ± 12 * |
| $\varepsilon_b$ /% | 888 ± 72 * | 885 ± 187 * | 461 ± 284 # | 185 ± 58 # | 54 ± 22 #* | 19 ± 5 #* |

**Figure S6.** Representative stress ($\sigma$)-strain ($\varepsilon$) curves (A-B) and tensile properties (C) of dry wet spun fibres made of either functionalised or native PVA. (A): F-TB (—), FA-TB-MA* (– · –), and FH-TB-MA* (···). (B): PVA10 (—), PVA12 (—) and PVA15 (···). (C): Calculated tensile modulus ($E$), maximal stress ($\sigma_{max}$) and strain at break ($\varepsilon_b$). Data are presented as mean ± standard deviation (n=5). # $p < 0.05$ vs PVA12; * $p < 0.05$ vs F-TB.

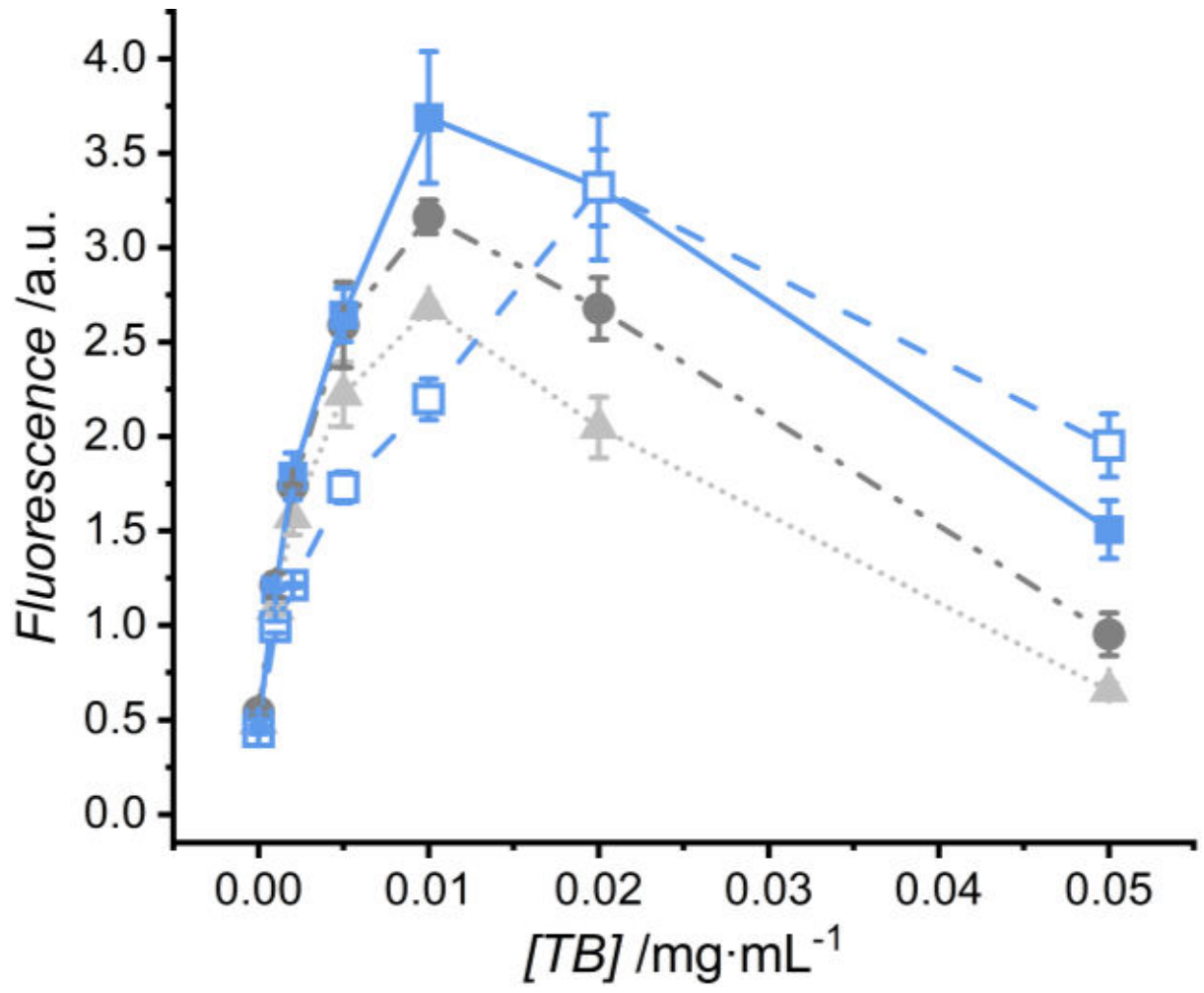

**Figure S7.** ROS assay fluorescence intensity measured on TB-supplemented PBS solution controls (n=3) following work light exposure for 10 min (···▲···), 20 min (—·●·—), 30 min (—■—) and 60 min (--□--). Data are presented as mean ± standard deviation (n=3).

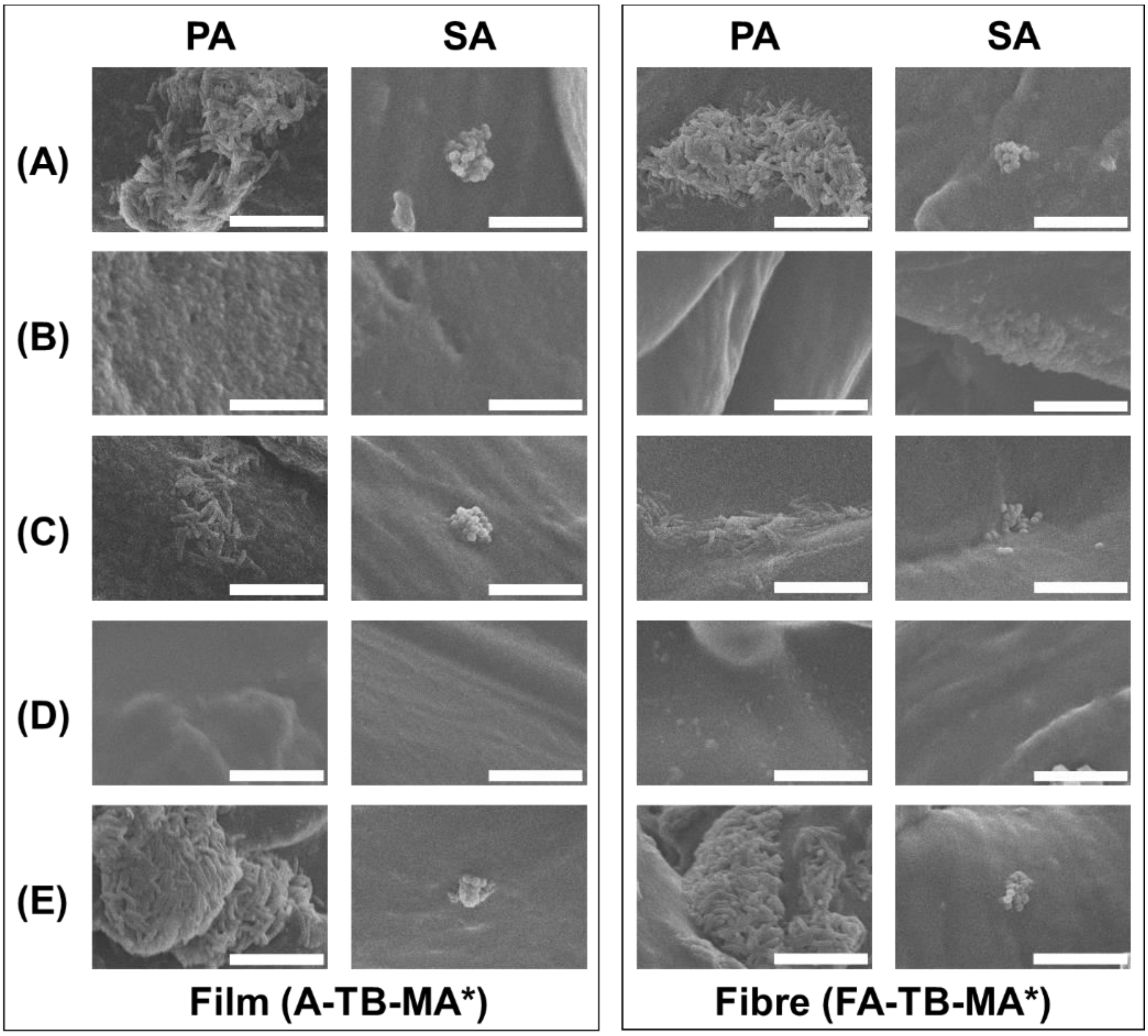

**Figure S8.** SEM images of either P. aeruginosa (PA) or S. aureus (SA) captured on samples of either film A-TB-MA* or fibre FA-TB-MA* prior to (A) and following (B-E) light irradiation. (B): 30-min irradiated samples; (C): 30-min dark controls; (D): 60-min irradiated samples; (E): 60-min dark controls. Scale bars: 10 μm.

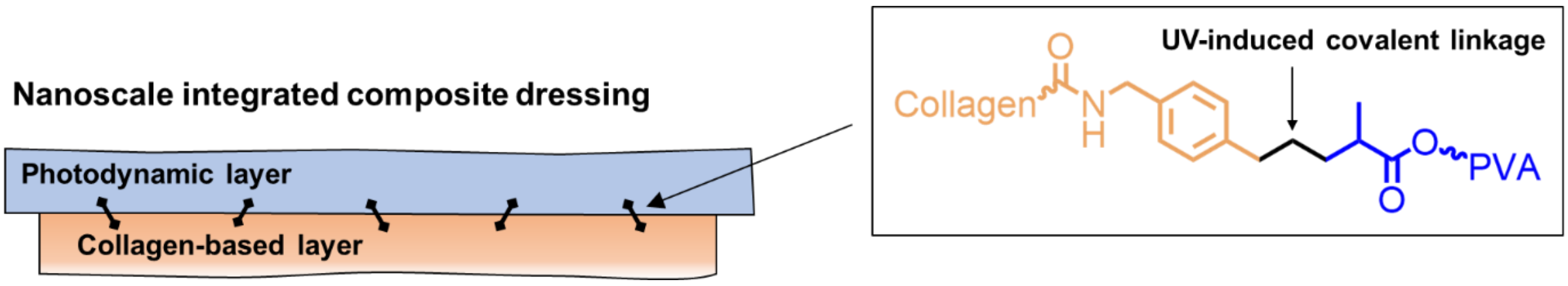

**Figure S9.** Schematic of the nanoscale integrated composite dressing featuring a collagen-based wound contact layer, a photodynamic antimicrobial layer and interface UV-induced covalent linkages.